\newcount\Ecount \Ecount=0
\def\Eeqn#1{\global\advance\Ecount by 1
  \global\xdef#1{\relax\the\Ecount} \eqno{\Edis#1}}
\def\Edis#1{\hbox{(D--#1)}}
\newcount\Dcount \Dcount=0
\def\Deqn#1{\global\advance\Dcount by 1
  \global\xdef#1{\relax\the\Dcount} \eqno{\Ddis#1}}
\def\Ddis#1{\hbox{(E--#1)}}
\newcount\Ccount \Ccount=0
\def\Ceqn#1{\global\advance\Ccount by 1
  \global\xdef#1{\relax\the\Ccount} \eqno{\Cdis#1}}
\def\Cdis#1{\hbox{(C--#1)}}
\newcount\Bcount \Bcount=0
\def\Beqn#1{\global\advance\Bcount by 1
  \global\xdef#1{\relax\the\Bcount} \eqno{\Bdis#1}}
\def\Bdis#1{\hbox{(B--#1)}}
\newcount\Acount \Acount=0
\def\Aeqn#1{\global\advance\Acount by 1
  \global\xdef#1{\relax\the\Acount} \eqno{\Adis#1}}
\def\Adis#1{\hbox{(A--#1)}}
\newcount\Fcount \Fcount=0
\def\Figref#1{\global\advance\Fcount by 1
  \global\xdef#1{\relax\the\Fcount} }
\def\Figno#1{Fig.~#1} 

\font\fourteenbf=cmbx10 scaled\magstep2
\font\twelvebf=cmbx10 scaled\magstep1

%
%
\raggedbottom
\baselineskip=16pt 
\overfullrule=0pt
\def\frac#1/#2{\leavevmode\kern.1em
 \raise.5ex\hbox{\the\scriptfont0 #1}\kern-.1em
 /\kern-.15em\lower.25ex\hbox{\the\scriptfont0 #2}}
\def\approx{\simeq}
%
%
\catcode`@=11
\newcount\chapternumber      \chapternumber=0
\newcount\sectionnumber      \sectionnumber=0
\newcount\equanumber         \equanumber=0
\let\chapterlabel=0
\newtoks\chapterstyle        \chapterstyle={\Number}
\newskip\chapterskip         \chapterskip=\bigskipamount
\newskip\sectionskip         \sectionskip=\medskipamount
\newskip\headskip            \headskip=8pt plus 3pt minus 3pt
\newdimen\chapterminspace    \chapterminspace=15pc
\newdimen\sectionminspace    \sectionminspace=10pc
\newdimen\referenceminspace  \referenceminspace=25pc
\def\chapterreset{\global\advance\chapternumber by 1
   \ifnum\the\equanumber<0 \else\global\equanumber=0\fi
   \sectionnumber=0 \makel@bel}
\def\makel@bel{\xdef\chapterlabel{%
\the\chapterstyle{\the\chapternumber}.}}
\def\sectionlabel{\number\sectionnumber \quad }
\def\unnumberedchapters{\let\makel@bel=\relax \let\chapterlabel=\relax
\let\sectionlabel=\relax \equanumber=-1 }
\def\eqname#1{\relax \ifnum\the\equanumber<0
     \xdef#1{{\rm(\number-\equanumber)}}\global\advance\equanumber by -1
    \else \global\advance\equanumber by 1
      \xdef#1{{\rm(\chapterlabel \number\equanumber)}} \fi}

\def\eqn#1{\eqno\eqname{#1}#1}

\def\eqinsert#1{\noalign{\dimen@=\prevdepth \nointerlineskip
   \setbox0=\hbox to\displaywidth{\hfil #1}
   \vbox to 0pt{\vss\hbox{$\!\box0\!$}\kern-0.5\baselineskip}
   \prevdepth=\dimen@}}
%

%

%

%
%
\newcount\fcount \fcount=0
\def\ref#1{\global\advance\fcount by 1 
  \global\xdef#1{\relax\the\fcount}}
\def\refs{\noindent \hangindent=5ex\hangafter=1}
\def\refsq{\noindent \hangindent=8ex\hangafter=1 \hskip 4ex}

%

%
%
\def\today{\number\year\space \ifcase\month\or 	January\or February\or 
	March\or April\or May\or June\or July\or August\or September\or
	October\or November\or December\fi\space \number\day}
%
%
\def\spose#1{\hbox to 0pt{#1\hss}}
\def\simlt{\mathrel{\spose{\lower 3pt\hbox{$\mathchar"218$}}
     \raise 2.0pt\hbox{$\mathchar"13C$}}}
\def\simgt{\mathrel{\spose{\lower 3pt\hbox{$\mathchar"218$}}
     \raise 2.0pt\hbox{$\mathchar"13E$}}}
\def\simpropto{\mathrel{\spose{\lower 3pt\hbox{$\mathchar"218$}}
     \raise 2.0pt\hbox{$\propto$}}}
%
%
\newif\ifmathmode
\def\mathflag#1${\mathmodetrue#1\mathmodefalse$}
\everymath{\mathflag}
\def\displayflag#1$${\mathmodetrue#1\mathmodefalse$$}
\everydisplay{\displayflag}
\mathmodefalse

\font\tenmmib=cmmib10
\font\sevenmmib=cmmib10
\font\fivemmib=cmmib10
\newfam\bmitfam\textfont\bmitfam=\tenmmib 
\scriptfont\bmitfam=\sevenmmib\scriptscriptfont\bmitfam=\fivemmib
\def\fixfam#1{\ifmathmode
                 {\ifnum\fam=\bffam
                    {\fam\bmitfam#1}\else
                    {\fam1#1}\fi}\else
                 {\ifnum\fam=\bffam
                    {$\fam\bmitfam#1$}\else
                    {$\fam1#1$}\fi}\fi}

\def\alpha{\fixfam{\mathchar"710B}}
\def\beta{\fixfam{\mathchar"710C}}
\def\gamma{\fixfam{\mathchar"710D}}
\def\delta{\fixfam{\mathchar"710E}}
\def\epsilon{\fixfam{\mathchar"710F}}
\def\zeta{\fixfam{\mathchar"7110}}
\def\eta{\fixfam{\mathchar"7111}}
\def\theta{\fixfam{\mathchar"7112}}
\def\iota{\fixfam{\mathchar"7113}}
\def\kappa{\fixfam{\mathchar"7114}}
\def\lambda{\fixfam{\mathchar"7115}}
\def\mu{\fixfam{\mathchar"7116}}
\def\nu{\fixfam{\mathchar"7117}}
\def\xi{\fixfam{\mathchar"7118}}
\def\pi{\fixfam{\mathchar"7119}}
\def\rho{\fixfam{\mathchar"711A}}
\def\sigma{\fixfam{\mathchar"711B}}
\def\tau{\fixfam{\mathchar"711C}}
\def\upsilon{\fixfam{\mathchar"711D}}
\def\phi{\fixfam{\mathchar"711E}}
\def\chi{\fixfam{\mathchar"711F}}
\def\psi{\fixfam{\mathchar"7120}}
\def\omega{\fixfam{\mathchar"7121}}
\def\varepsilon{\fixfam{\mathchar"7122}}
\def\vartheta{\fixfam{\mathchar"7123}}
\def\varpi{\fixfam{\mathchar"7124}}
\def\varrho{\fixfam{\mathchar"7125}}
\def\varsigma{\fixfam{\mathchar"7126}}
\def\varphi{\fixfam{\mathchar"7127}}

\unnumberedchapters
 

 at 14.4truept
\font\namefont=cmr12 
\font\addrfont=cmti12 

\newbox\abstr
\def\abstract#1{\setbox\abstr=
    \vbox{\hsize 5.0truein{\par\noindent#1}}
    \centerline{ABSTRACT} \vskip12pt 
    \hbox to \hsize{\hfill\box\abstr\hfill}}

\def\today{\ifcase\month\or
        January\or February\or March\or April\or May\or June\or
        July\or August\or September\or October\or November\or December\fi
        \space\number\day, \number\year}

\def\author#1{{\namefont\centerline{#1}}}
\def\addr#1{{\addrfont\centerline{#1}}}

\def\bk {{\bf k}}
\def\bg {{\bf \gamma}}

\def\ie {{\it i.e.}}
\def\eg {{\it e.g.}} 
\def\etal {{\it et al.}}
\def\Vg {{\cal V}_\gamma}
\def\Vb {{\cal V}_b}
\def\Dg {{\cal D}_\gamma}
\def\Db {{\cal D}_b}

\def\section#1{{\goodbreak\bigskip\noindent\fourteenbf#1}}
\def\subsection#1{{\smallskip\noindent\bf#1}}

{\nopagenumbers
\vsize=9 truein
\hsize=6.5 truein
\raggedbottom
\baselineskip=18pt
\hfill IAS/SNS-AST-95/42, CfPA-TH-95-18, UTAP-212 
 
\hfill 
\vskip2truecm
\centerline {
 \twelvebf  
{\fourteenbf S}MALL 
{\fourteenbf S}CALE 
{\fourteenbf C}OSMOLOGICAL 
{\fourteenbf P}ERTURBATIONS: }


\centerline {
\twelvebf
{\fourteenbf A}N
{\fourteenbf A}NALYTIC
{\fourteenbf A}PPROACH
\footnote{$^{\rm\dag}$}{\rm\negthinspace\negthinspace\negthinspace
Submitted to ApJ, revised \today} 
	    }
\nobreak
\baselineskip=15pt
  \vskip 0.5truecm
  \author{Wayne Hu$^{1}$ and Naoshi Sugiyama$^{2}$}
  \smallskip
  \addr{$^{1}$Institute for Advanced Study}
  \addr{School of Natural Sciences}
  \addr{Princeton, NJ 08540}
  \smallskip
  \addr{$^{2}$Department of Physics, Faculty of Science}
  \addr{The University of Tokyo, Tokyo 113, Japan}
  \bigskip
  \bigskip	 
\noindent{\rm 
Through analytic techniques verified by numerical calculations, 
we establish general relations between the matter and cosmic
microwave background (CMB) power spectra and their dependence on
cosmological parameters on small scales. Fluctuations in the CMB,
baryons, cold dark matter (CDM), and neutrinos receive a boost at horizon
crossing.  Baryon drag on the photons
causes alternating acoustic peak heights in the 
CMB and is uncovered
in its bare form under the photon diffusion scale.
Decoupling of the photons at last scattering and 
of the baryons at the end of the Compton drag epoch, freezes the
diffusion-damped acoustic oscillations into the CMB and matter power spectra at different scales.
We determine the dependence of the respective acoustic
amplitudes and damping lengths on fundamental cosmological parameters.
The baryonic oscillations, enhanced by the velocity overshoot effect, 
compete with CDM fluctuations in the present matter power spectrum.
We present new {\it exact} analytic solutions for the
cold dark matter fluctuations in the presence of a growth-inhibiting
radiation {\it and} baryon background.  Combined with the acoustic
contributions and baryonic infall into CDM potential wells, this
provides a highly accurate analytic form of the small-scale 
transfer function in the general case.

\smallskip\noindent
{\it Subject headings:} cosmic microwave background --
large-scale structure of universe -- cosmology:theory
}
}
\bigskip
\vfill 
\noindent whu@sns.ias.edu

\noindent sugiyama@yayoi.phys.s.u-tokyo.ac.jp
\eject

\Figref\cmbtime
\Figref\dampvis
\Figref\rad
\Figref\bartime
\Figref\trans
\Figref\cdmtime
\Figref\ftwo
\Figref\rpsi
\Figref\rpsils
\Figref\cdmexact
\Figref\velov
\Figref\transcdm

\input epsf.tex

\font\ninesl=cmsl9 
{ \ninesl
\hskip 3.5truein \baselineskip= 12pt
The sage sets aside obfuscation

\hskip 3.5truein
And is indifferent to baseness and honor

\hskip 3.5truein
The mass of men are all hustle-bustle

\hskip 3.5truein
The sage is slow and simple

\hskip 3.5truein
He combines myriad years

\hskip 3.5truein
Into a single purity

\hskip 3.5truein
Thus does he treat the myriad things

\hskip 3.5truein
And thereby gathers them together

\hskip 4.5truein 
--Chuang-tzu, 2 }

\bigskip

\section{I. Introduction}

Small scale fluctuations in the cosmic microwave background (CMB)
and the matter density
provide a unique opportunity to probe structure
formation in the universe.  As CMB anisotropy experiments reach toward
smaller and smaller angles, the region of overlap with large scale structure
measurements will increase dramatically.  
Since the CMB and matter power spectra 
encode information at very different epochs in the formation
of structure, comparison of the two will provide a powerful 
consistency test for competing scenarios.  Unfortunately, the
simple relation between the two power spectrum at large scales
(Sachs \& Wolfe 1967) does not
hold at small scales.  To establish the general relation, we
need to employ a more complete analysis of how gravitational 
instability affects the matter and CMB together.

In previous papers  (Hu \& Sugiyama 1995a,b, hereafter HSa,b), 
we developed a conceptually simple 
analytic description of CMB anisotropy formation.  The central
advance over previous analytic works (\eg\ Doroshkevich, Zel'dovich 
\& Sunyaev 1978) involved 
the influence of gravitational potential wells, established 
by the decoupled matter, on the acoustic oscillations in the photon-baryon
fluid. 
In this paper, we refine the analysis 
for scales much smaller than the horizon
at last scattering as is relevant for large scale structure calculations. 
We furthermore 
extend it to encompass the photon-baryon backreaction
on the evolution of the cold dark matter and gravitational potential 
as well as baryonic decoupling and evolution. 
This makes it possible to extract the relation between CMB and matter
fluctuations established in the early universe. 

Analytic solutions in terms of 
{\it elementary} functions can be constructed in the small scale limit and 
serve to illuminate the physical processes involved in a model-independent
manner. 
Despite the simplicity of the final results, the study of small scale
fluctuations requires a rather technical exposition of perturbation
theory.  For this reason, we divide this paper into two components.
The main text discusses results drawn from a series of appendices 
and illustrates the corresponding principles in the familiar 
context of adiabatic cold dark matter (CDM), adiabatic baryonic dark matter
(BDM), and isocurvature BDM scenarios for structure formation. 

We begin in \S 2 with a discussion of the 
central approximations employed, \ie\ the tight coupling limit
for the photons and baryons and the external potential representation
for metric fluctuations.  Details of these methods and the justification
of their use can be found in Appendix A.  As established in Appendix B, 
all fluctuations are given a boost at horizon crossing due to the driving 
effects of gravitational infall and dilation.   Since the gravitational
potential subsequently decays, the driving effects disappear well after
horizon crossing, leaving fluctuations to evolve in their natural 
source-free modes.  For the photon-baryon system, these are acoustic
oscillations whose zero point is displaced by baryon drag.  As described
in \S 3 and \S 4, these oscillations  are 
frozen into the photon and baryon spectra at last scattering and
the end of the Compton drag epoch respectively. 
Since these epochs are not equal, photon diffusion (Silk 1968)
sets a different
damping scale in the CMB and the baryons.  This process  is
described by the acoustic visibility functions introduced in Appendix C.

The baryonic oscillations
may be hidden in CDM models by larger cold dark matter fluctuations.
These are discussed in \S 5.  The baryon and radiation {\it background} also
suppresses CDM growth before the end of the Compton drag epoch.  
Appendix D treats this source-free evolution of the CDM component
analytically. 
After the end of the Compton drag epoch, the CDM fluctuations provide 
potential wells into which the baryons fall.  From the
combined analysis of CMB, baryon and CDM fluctuations, we obtain 
accurate analytic expressions for matter and photon transfer functions 
in the small scale limit. This greatly improves upon the fitting functions
for the matter power spectrum 
in the literature (Bardeen \etal\ 1986, hereafter BBKS, 
Peacock \& Dodds 1994, Sugiyama 1995)
in the case of a significant baryon fraction.
Our form is constructed out of fits from Appendix E for parameters
that depend on the detailed physics of recombination, \ie\
the last scattering epoch, Compton drag epoch, photon
damping length, and Silk damping length.

\section{2. Acoustic Approximation}

We previously
developed an 
analytic description of acoustic oscillations in the photons and baryons
(HSa,HSb).  
As discussed in more detail in Appendix A, this approach
is based on two simplifying assumptions: that a perturbative expansion
in the Compton scattering time between photons and electrons 
can be extended up to recombination and
that the gravitational potential may be considered as an external
field. 
Combined these two simplifications imply that photon pressure resists
compression by
infall into potential wells and sets up acoustic oscillations
in the photon-baryon fluid which then damp by photon diffusion.  However,
 both of these approximations would seem to be problematic for 
small-scale fluctuations.  Here, the optical depth through a wavelength
of the fluctuation at last scattering is small implying 
weak rather than tight coupling (see \eg\ Kaiser 1984,
Hu \& White 1995).  
This issue is addressed more fully in Appendix C [below eqn.~(C-3)] 
where it is 
shown that only acoustic contributions from the 
tight coupling regime are visible through the last scattering surface,  
\ie\ those from a time {\it slightly} earlier than last scattering 
when the optical depth to 
Compton scattering was still high.  


On the other hand, the second issue 
poses a computational problem.  Inside the horizon
in the radiation 
dominated epoch, the acoustic oscillations in the photon-baryon density 
feed back into the evolution 
of the gravitational 
potential through the generalized
Poisson equation.  Hu \& White (1996) take the approach of including
the self-gravity effects consistently by separating them from the
truly external potentials. 
For many cases, a simpler approach suffices.  By considering the
self-gravity of the fluid as contributing to 
an external potential, we retain the
conceptual simplicity of HSa and HSb.  
Although a consistent evolution of the potential would seem 
to require a full numerical solution of the coupled equations, it is
in practice not generally necessary.   
The key point is that the feedback effect on the potentials is not arbitrary.
Since photon pressure prevents the growth of photon-baryon fluctuations and 
collisionless damping eliminates neutrino contributions, the potential 
decreases to zero after horizon crossing in the radiation dominated
epoch.   We shall examine how this affects the acoustic oscillations
more carefully in \S 3.

Due to Compton drag from the coupling,
the baryons density fluctuations follow
the photons as $\dot \delta_b = {3 \over 4} 
\dot \delta_\gamma =  3\dot \Theta_0$ or $\delta_b = 3 \Theta_0 + S$ 
and exhibit acoustic oscillations as well.
Here overdots represent derivatives with respect to conformal time,
$\Theta_0$ is the isotropic
temperature perturbation
in Newtonian gauge, and $S$ is the constant photon-baryon entropy.  
On the other hand, the cold dark matter, if present, is decoupled from
photons and suffers only the gravitational effects of the oscillating
radiation.  Again, since the potential created by the radiation merely
oscillates and damps away after horizon crossing, the CDM 
density perturbations $\delta_c$ experience
a kick at horizon crossing only to settle into the pure logarithmically
growing mode in the radiation dominated universe.

\topinsert
\centerline{
\epsfxsize=3.5in \epsfbox{cmbtime.epsf}
}

\baselineskip=12truept \leftskip=3truepc \rightskip=3truepc 
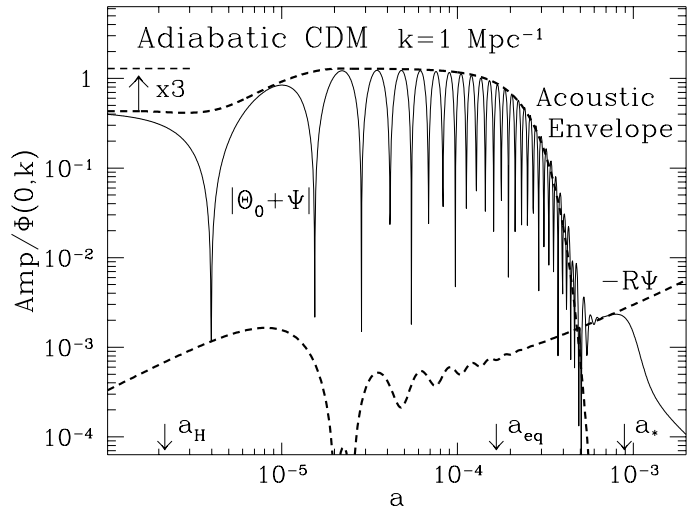
\noindent {\bf \Figno\cmbtime.} Acoustic oscillations receive a 
boost at horizon crossing $a_H$ due driving from gravitational
potential decay.  
The perturbations then settle into a pure acoustic mode and are
subsequently damped
by photon diffusion.  
Together the potential driving and diffusion damping effects form the
acoustic envelope.  After diffusion damping has destroyed the acoustic
oscillations, the underlying baryon drag effect becomes apparent.  
Since $\Psi$ is here constant for $a \gg a_{eq}$, $|R\Psi|$ increases
with time.
This contribution itself is damped at last scattering $a_*$ by cancellation.
Well after last scattering $a \gg a_*$, isotropic perturbations 
collisionlessly damp creating angular fluctuations in the CMB.
The model here is adiabatic CDM, $\Omega_0=1$, $h=0.5$, $\Omega_b=0.05$.
\endinsert
\section{3. Photons Fluctuations}

\subsection{3.1 Driving Effects}

Before recombination, the photon-baryon system behaves as a damped,
forced oscillator.  As discussed in \S 2, the self-gravity of the
acoustic oscillations contributes to the gravitational force on the
oscillator at horizon crossing if the photons or baryons contribute
significantly to the density of the universe at that epoch.  
Even though the detailed behavior 
at horizon crossing in the radiation dominated epoch
is complicated by feedback effects, the end result
is extremely simple: fluctuations experience a boost proportional
to the potential at horizon crossing.  
As the first compression of the fluid is halted by photon pressure,
the self-gravitational potential decays.  This leaves the oscillator
in a highly compressed state and enhances the amplitude of the subsequent
oscillation.
For adiabatic initial conditions,
the amplitude of the 
potential at horizon crossing is essentially the initial curvature fluctuation.
For isocurvature initial conditions, it is simply related to the initial entropy
perturbations.  

We quantify these statements in Appendix B and calculate
the exact boost factor [see eqns.~(B-5),(B-6)]. 
It represents 
an enhancement in temperature fluctuations of a factor of three 
over its initial value in the adiabatic case or more importantly a factor of
$5/(1+ {4 \over 15} f_\nu)$ over 
the gravitational redshift induced
large-scale fluctuations in a scale-invariant model (Sachs \& Wolfe 1967).  
Here the fraction of
the radiation energy density contributed by the neutrinos is $f_\nu = 
\rho_\nu / (\rho_\nu + \rho_\gamma) \approx 0.405$ 
for the standard thermal history with three massless species.
Its presence accounts for neutrino anisotropic
stress which provides 
the relation between the gravitational potential $\Psi$ 
and the 
curvature fluctuation $\Phi$ [see eqn.~(A-5)].
Note that neutrino temperature fluctuations 
are boosted by a similar factor (Hu \etal\ 1995).

Three features are worth emphasizing:
\smallskip
\refs (1) Potential decay {\it enhances} temperature perturbations.

\refs (2) Enhancement only occurs for scales that cross the horizon 
{\it before} equality.

\refs (3) It occurs {\it at} horizon crossing causing phase-coherent
oscillations in the wavenumber $k$.

\smallskip
\noindent
Together the first two points imply that as a function of $k$, the
acoustic amplitude will increase through $k_{eq}$, the horizon
crossing wavenumber at equality, forming a potential envelope.
By delaying equality, one moves the transitional regime to larger
scales and enhances the lower $k$ oscillations.
This explains the increasing prominence
of the first few acoustic peaks as 
the matter content $\Omega_0 h^2$ is lowered or the number of
relativistic species raised in adiabatic
models.  It is interesting to note that the effect of equality on
the CMB and matter power spectrum are anticorrelated providing a 
powerful consistency test for its redshift $z_{eq}$.  
As we shall see in \S 3.2 though, a complication arises since
high-$k$ oscillations are damped by photon diffusion.

The third point follows 
because the gravitational driving term is effective at 
sound horizon crossing
$\eta \sim 1/kc_s$, thus mimicing a driving frequency of $\tilde \omega
\sim kc_s$.  Here $\eta = \int dt/a$ is the conformal time and
we take $c=1$ throughout.
Since the natural frequency of the oscillation
is related to the sound speed $c_s$ as $\omega = kc_s$,
the two scale in the same manner.
Thus the horizon crossing effect is timed to produce 
a coherent oscillation in $k$.  Specifically, adiabatic and isocurvature
initial conditions yield cosine and sine temperature oscillations 
respectively.

\subsection{3.2. Damping Effects}

The photons and baryons are not in fact perfectly coupled leading
to diffusion damping.
The coupling strength is quantified by the Compton 
optical depth $\tau$.  Correspondingly, the  
mean free path of the photons in the
baryons is given by $\dot \tau^{-1}$.  Explicitly,
$\dot \tau = x_e n_e \sigma_T a$ where 
$x_e$ is the electron ionization fraction, $n_e$ is the electron
number density, and $\sigma_T$ is the Thomson cross section. 
As the photons random walk across a wavelength of the perturbation,
temperature fluctuations collisionally damp. More specifically, diffusion
generates viscosity or anisotropic stress
in the photon-baryon fluid and
causes heat conduction across the wavelength (Weinberg 1972).
Both of these processes damp fluctuations. To order of magnitude,
the damping length is the random walk distance
$\sqrt{\eta/\dot\tau}$.  It exceeds the wavelength of a fluctuation
when the optical depth through a single wavelength $\dot \tau / k = k\eta$.  
If the diffusion scale is well under the horizon, $k\eta \gg 1$ 
so that $\dot \tau/k$ is still high at crossover. 
The photons and baryons are thus still strongly
coupled and the damping may be calculated under the tight coupling
approximation. 

A quantitative
treatment of diffusion damping is given in Appendix A.  
Subtle effects such as the angular and polarization
dependence of Compton scattering slightly enhance the generation of viscosity
and thus damping in the radiation dominated universe (Kaiser 1983). 
The end result of the calculation is
a wavenumber $k_D(\eta)$ by which acoustic fluctuations
are damped as $\exp[-(k/k_D)^2]$ [see eqn.~(A-14)].  
Combined with the potential envelope from the
horizon crossing boost of \S 3.1, 
this completes the acoustic envelope shown in \Figno\cmbtime.
Remaining after diffusion damping is
the acoustic offset of $\Theta_0 + \Psi = -R\Psi$, where 
$R= 3\rho_b/4\rho_\gamma = 31.5 \Omega_b h^2 \Theta_{2.7}^{-4} (z/10^3)^{-1}$.  
Here $\Theta_{2.7} = T_0/2.7$K is the scaled present temperature of the
CMB, $\Omega_b$ is the fraction of critical density contributed by 
the baryons, and the Hubble constant is 
$H_0 = 100h$ km s$^{-1}$ Mpc$^{-1}$.
We add the Newtonian potential $\Psi$ to 
the temperature perturbation $\Theta_0$ 
to remove the effect of gravitational redshift
on the photons (see Appendix A).  
The $R\Psi$ term represents the baryon drag effect on
the photons and is analogous to the Silk mechanism  (Silk 1968) with the roles
of the photons and baryons interchanged.  
Infalling baryons drag the photons into
potential wells leading to a displacement of the
zero point of the acoustic oscillation (HSa).  Thus it is responsible
for the alternating peak heights and amplitude enhancement 
of intermediate scale acoustic oscillations (see Appendix A, Fig.~8a).  
The zero-point
shift remains even after diffusion damping has eliminated the oscillations
themselves. 

\subsection{3.3. Decoupling}

As the total optical
depth to the present drops below unity $\tau(z_*)=1$, 
last scattering of the CMB photons 
freezes the acoustic oscillations into
the spectrum.  
The optical depth drops rapidly 
as neutral hydrogen forms 
so that last scattering and recombination approximately coincide
in the absence of subsequent reionization.
This epoch is nearly independent of cosmological parameters such 
as the matter and baryon content $\Omega_0 h^2$ and $\Omega_b h^2$ 
(Peebles 1968, Jones \& Wyse 1985). 
However variations at the $10\%$ level do occur across the full range
of parameters.  In Appendix E [eqn.~(E-1)], 
we present an extremely accurate
analytic fit to the last scattering epoch.  

The phase of the acoustic oscillation is frozen in at
the value $k r_s(\eta_*)$, where  (HSa)
$$
r_s = 2  {\sqrt{3} \over 3}  (\Omega_0 H_0^2)^{-1/2} \sqrt{ {a_{eq} \over
	R_{eq} } } \ln { \sqrt{ 1 + R} + \sqrt{ R + R_{eq}} 
	\over 1 + \sqrt{R_{eq}} },
\eqn\eqnSound
$$
is the sound horizon with $R_{eq} = R(a_{eq})$
and $a_{eq} = 2.35 \times 10^{-5} (\Omega_0 h^2)^{-1} (1-f_\nu)^{-1}
\Theta_{2.7}^4$.
Its
variation with $k$ produces an oscillatory pattern in the
CMB temperature with extrema at scales (HSb)
$$
k_{\gamma j} r_s(\eta_*) = \cases { j\pi, & adiabatic \cr
			   (j - 1/2)\pi. & isocurvature \cr }  
\eqn\eqnGammaPeak
$$
The physical scale of the peaks $k_{\gamma j}$ is projected onto 
an angular scale on the sky 
and provides a sensitive angular diameter distance 
test for curvature in the universe.  

\topinsert
\centerline{ \hskip -0.5 truecm
\epsfxsize=3in \epsfbox{dampvisa.epsf} \hskip 0.5truecm
\epsfxsize=3in \epsfbox{dampvisb.epsf}
}

\baselineskip=12truept \leftskip=3truepc \rightskip=3truepc 
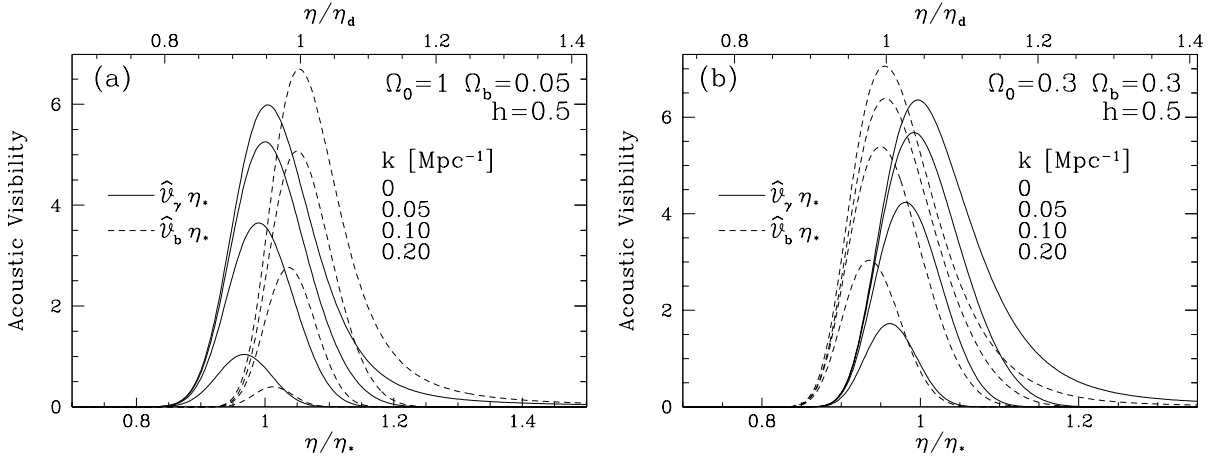
\noindent {\bf \Figno\dampvis.} Visibility functions for the
photons and baryons.  
In the undamped $k\rightarrow 0$ limit, the photon acoustic 
visibility and the Compton visibility
are equivalent $\hat \Vg = \Vg$ and the baryon acoustic visibility
equals the drag visibility, $\hat\Vb =  \Vb$. 
If $\Omega_b h^2 \simlt 0.03$,
$\Vb$ peaks at later
times than $\Vg$, \ie\ $\eta_d > \eta_*$.
For small scales, the acoustic visibilities, which weight the 
acoustic contributions from the tight coupling regime, are increasingly 
suppressed at later times by damping 
and the width and amplitude of the acoustic visibilities
decrease as $k$ increases.  
\endinsert

To determine the amplitude of the fluctuations, 
this instantaneous decoupling approximation must be modified to
account for diffusion damping through recombination.
The differential visibility function $\Vg = \dot\tau e^{-\tau}$ expresses
the probability that a photon last scattered within $d\eta$ of $\eta$.  
This function describes how fluctuations at recombination are frozen
into the CMB.  Equally useful for our purposes is the acoustic
visibility function $\hat \Vg = \Vg \exp[-(k/k_D)^2]$.  By including
the intrinsic damping behavior of the fluctuations, it describes
how acoustic oscillations are frozen into the CMB (see \Figno\dampvis).  
Due to the growth of the diffusion length
through recombination, this function is weighted to slightly earlier 
times than $\Vg$.  At small scales, only the small fraction of
photons last which last scattered {\it before}
$z_*$, and hence in the tight coupling regime, 
retain acoustic fluctuations.  This leads to a sharp decrease
in acoustic fluctuations with $k$.  More specifically, the damping envelope
is given by (HSa)
$$
\Dg(k) = \int_0^{\eta_0} d\eta \hat \Vg(\eta,k) 
 = \int_0^{\eta_0} d\eta \Vg(\eta) e^{-[k/k_D(\eta)]^2},
\eqn\eqnGammaDamp
$$
\ie\ a near exponential damping in $k$.  Through the first decade
of the drop, it is well approximated by the form
$$
\Dg(k) \approx e^{-[k/k_{D\gamma}]^{m_\gamma}}.
\eqn\eqnGammaDampApp
$$
The damping angle is given by the same projection conversion as
the acoustic peaks $\ell_{D} = k_{D\gamma} r_\theta(\eta_*)$.
Accurate fitting functions for $k_{D\gamma}$ and $m_\gamma$ are given 
in Appendix E, eqns.~(E-4) (E-7).  
Note that $k_{D\gamma}$ is the wavenumber at which diffusion
suppresses the fluctuation by $e^{-1}$.  It is interesting to note that  
$k_{D\gamma}^{-1}$ is intimately related to the width of the Compton
visibility function.  This is because the thickness of the last scattering
surface is by definition the diffusion length at last scattering.
%
%
%
%
%
%
\topinsert
\centerline{
\epsfxsize=3.0in \epsfbox{rbdma.epsf}
}
\centerline{
\hskip -.5 truecm \epsfxsize=3.0in \epsfbox{rbdmi.epsf} 
\hskip .5 truecm
\epsfxsize=3.0in \epsfbox{rcdm.epsf}
}

\baselineskip=12truept \leftskip=3truepc \rightskip=3truepc 
\noindent {\bf \Figno\rad}.
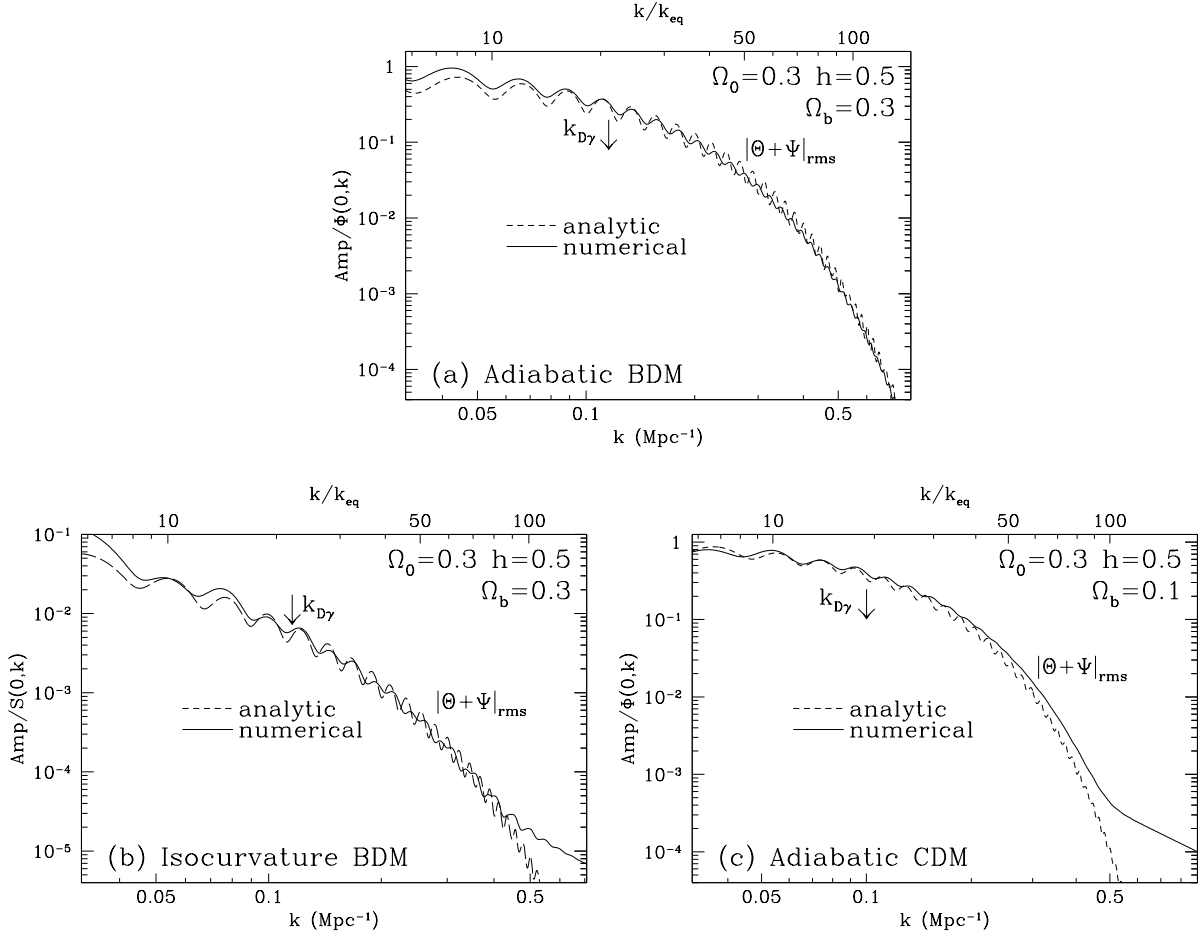
Photon transfer function. Plotted is the present rms photon temperature
fluctuation relative to the initial curvature fluctuation $\Phi(0,k)$ for
adiabatic fluctuations and the initial entropy fluctuation $S(0,k)$ for
isocurvature fluctuations assuming standard recombination.  
The intrinsic acoustic amplitude
is approximated by the analysis in Appendix B for scales well inside
the horizon at equality $k \gg k_{eq}$.  
The damping is well approximated by the tight coupling
expansion for high $\Omega_b h^2$ and is slightly overestimated for low
values.  Below the damping tail, the baryon drag offset clearly
appears in (b) and (c) where 
the gravitational potential is not dominated by acoustic density fluctuations.
An analytic treatment of this effect is given in Appendix A and C.
\endinsert
%
%
Combining the damping envelope with the intrinsic amplitude of acoustic oscillations
discussed in \S 2 and Appendix B, we obtain the transfer function 
at small scales.  This function represents the growth to the present
from the initial curvature or entropy perturbation.
In \Figno\rad, we plot examples for the adiabatic BDM,
isocurvature BDM, and adiabatic CDM models with standard recombination. 
Note that in more typical isocurvature BDM models (Peebles 1987a,b), 
reionization may occur soon after recombination.  
In this limit, 
the damping length continues to grow until it reaches the horizon
at the new last scattering surface and destroys all oscillations in
the photons.

The corresponding anisotropy 
can be obtained by choosing an initial curvature and entropy 
spectrum $\Phi(0,k)$ and $S(0,k)$ and
employing the free streaming solution (see HSa eq.~12)
for the  monopole and dipole contributions to the rms given in Appendix B,
eqn.~(B-7). 
Note that for low 
$\Omega_b h^2$ models, the damping length is somewhat overestimated
by the tight coupling approximation.  This is not surprising since as
$\Omega_b h^2 \rightarrow 0$, the photon mean free path approaches
the horizon.  In this case, the diffusion length passes the wavelength of the
fluctuation when the optical depth through a wavelength is near unity and
the tight coupling expansion of Appendix A breaks down.  The damping length
is {\it over}estimated because the photons essentially free stream across
the wavelength and do not suffer collisional damping.  A phenomenological
correction for this effect is given in Appendix E.

In summary, the general features of the CMB fluctuation spectrum
are
\smallskip
\refs (1) Acoustic peaks at harmonics of the angle the sound horizon
subtends at last scattering, sensitive to

\refsq (a) The curvature,

\refsq (b) The adiabatic or isocurvature nature of the initial
		fluctuations;

\smallskip
\refs (2) An acoustic envelope for the oscillation amplitude sensitive to 

\refsq (a) The matter-radiation ratio, \eg\ the adiabatic acoustic envelope
varies from ${1 \over 3}\Psi$ to ${5 \over 3}\Psi$ as one passes through
the equality scale,

\refsq (b) The baryon-photon ratio due to the a modulation of $R\Psi$
from baryon drag,

\refsq (c) The thermal history from the exponential diffusion-damping
tail, 

\smallskip

\noindent
though the exact nature will depend on details of the model.

\topinsert
\centerline{
\epsfxsize=3.5in \epsfbox{bartime.epsf}
}

\baselineskip=12truept \leftskip=3truepc \rightskip=3truepc 
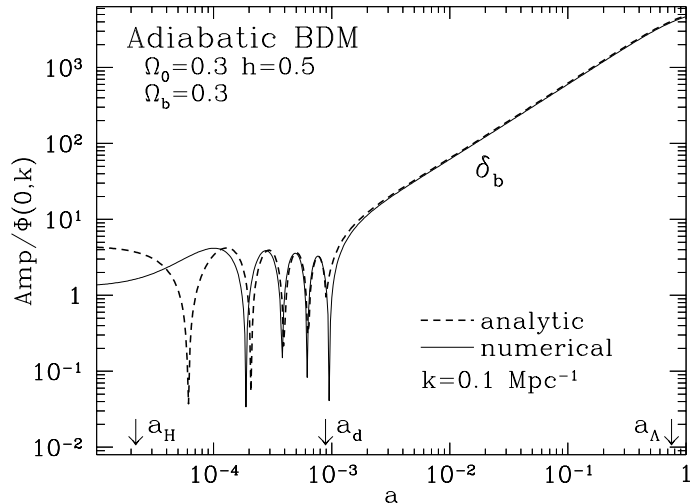
\noindent {\bf \Figno\bartime.} Baryon fluctuation 
time evolution.  The baryon
density fluctuation $\delta_b$ follows the photons before
the drag epoch $a_d$ yielding a simple oscillatory form
for $a_H \ll a \ll a_d$.  The Silk damping length is
given by the diffusion length at the drag epoch.  The portion of
the baryon fluctuations that enter the growing mode is dominated
by the velocity perturbation at  the drag epoch 
$a_d$ due to the velocity overshoot 
effect (see also \Figno\velov).  
Since a flat $\Lambda$ model is chosen here, the growth rate is
slowed by the rapid expansion for $a \simgt a_\Lambda = 
(\Omega_0/\Omega_\Lambda)^{1/3}$.
\endinsert
\section{4. Baryon Fluctuations}

In the tight coupling limit, the baryons participate in the acoustic
oscillations of the photons.  Near recombination, they 
decouple from the photons though not exactly at last scattering.
Scattering represents an exchange of momentum between the two fluids and 
seeks to equalize their bulk velocities $V_b$ and $V_\gamma$.  
However the two momentum densities $(\rho_\gamma + p_\gamma)
V_\gamma = { 4 \over 3} \rho_\gamma V_\gamma$ 
and $(\rho_b + p_b)V_b \approx \rho_b V_b$ are not equal.  
Thus momentum conservation requires that the rate of change of the baryon
velocity due to Compton drag is scaled by a factor of $R^{-1}= {4 \over 3}
\rho_\gamma / \rho_b$ compared with the photon case:  \ie\ 
$\propto \dot \tau_d = \tau/R$.   The explicit expression
for the baryon momentum conservation or Euler 
equation is given in Appendix A and is used in Appendix
C to make the qualitative statements here rigorous.  
The form of the coupling suggests that we can define a drag depth 
$\tau_d(\eta_d) = \int_{\eta_d}^{\eta_0} \dot \tau_d d\eta$.  Below drag depth 
$\tau_d(\eta_d) = 1$, the baryons dynamically decouple from the photons.
For the standard recombination scenario, this occurs near recombination 
but at a different value than last scattering.  Analytic fitting formulae
for $z_d$ are given in Appendix E, equation (E-2). 
Since recombination occurs around $z = 10^3$ for all models, the end of
the drag epoch precedes last scattering if $\Omega_b h^2 \simgt 0.03$.

\topinsert
\centerline{
\epsfxsize=3.0in \epsfbox{tbdma.epsf}
}
\centerline{
\hskip -.5 truecm \epsfxsize=3.0in \epsfbox{tbdmi.epsf} 
\hskip .5 truecm
\epsfxsize=3.0in \epsfbox{tcdm.epsf}
}

\baselineskip=12truept \leftskip=3truepc \rightskip=3truepc 
\noindent {\bf \Figno\trans}.
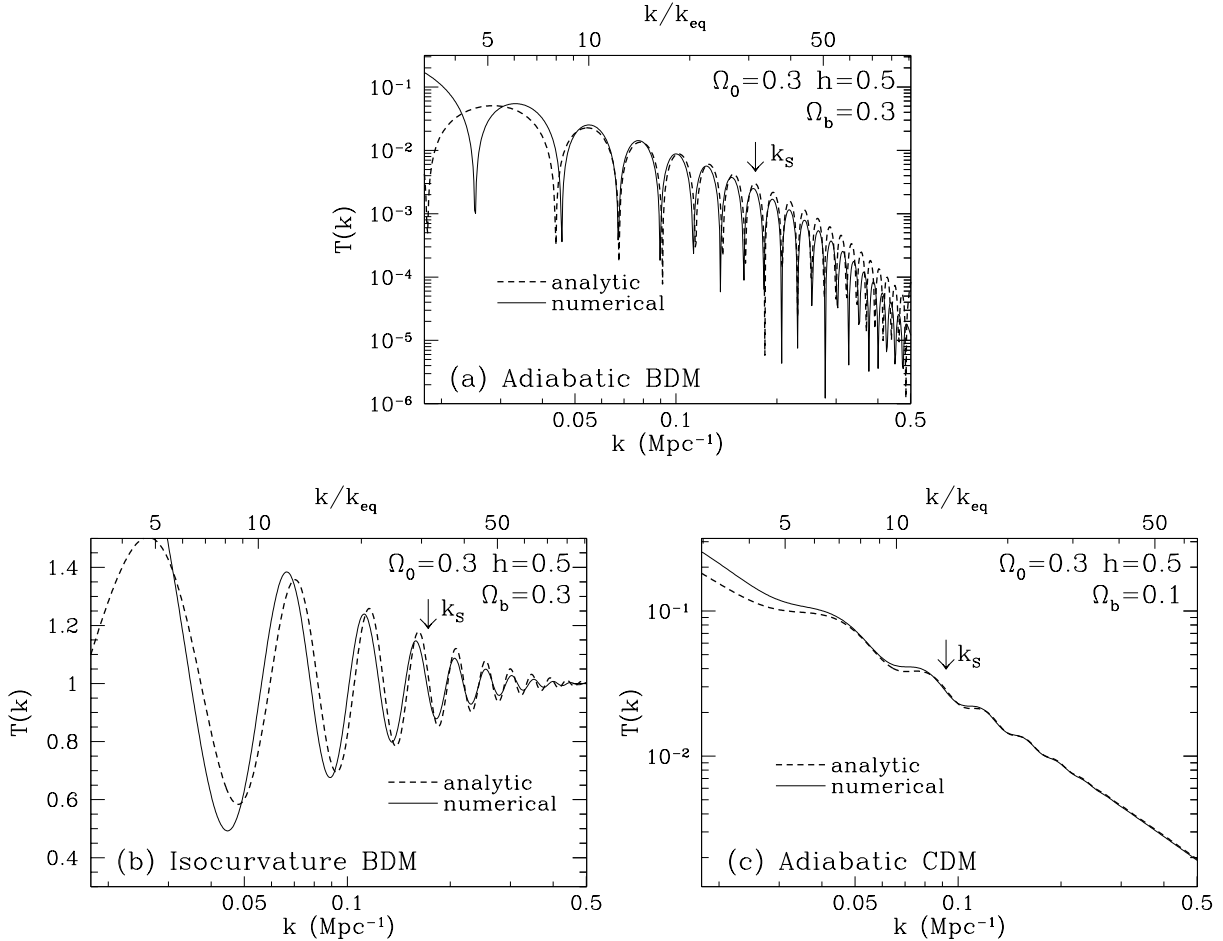
Matter transfer function.  The analytic estimates of the intrinsic
acoustic amplitude is a good approximation for $k \gg k_{eq}$.  
The Silk damping scale is adequately approximated although
its value is underestimated by $\sim 10\%$.  For isocurvature
BDM and adiabatic CDM, the acoustic contributions do not dominate
the small scale fluctuations.  We have added in the contributions
from the initial entropy fluctuations and the cold dark matter
potentials according to the analytic treatment of Appendix E. 
\endinsert
The acoustic fluctuations in the baryons are frozen in at 
the drag epoch rather than at last scattering.  Furthermore, unlike 
for the CMB, it is not the acoustic density fluctuation that forms
peaks in observable spectrum today but rather the acoustic velocity.  
This is because the baryon fluctuations continue
to evolve.
In Appendix C, we give the exact matching conditions at $z_d$ onto the growing 
and decaying modes of pressureless perturbation theory.  This
yields an accurate description for the subsequent evolution of the
baryonic fluctuations in the presence of a background radiation
energy density and cold dark matter.  Qualitatively, the acoustic velocity
at the drag epoch dominates over the acoustic density for 
the growing mode of fluctuations due to the velocity overshoot effect 
(Sunyaev \& Zel'dovich 1970, Press \& Vishniac 1980).
The former moves matter
and produces clumping in the baryon density.  Since expansion damps peculiar 
velocities, this lasts for approximately an expansion time $\eta_d$.
Thus only
 scales smaller than the horizon $k\eta_d \gg 1$ experience a boost due 
to the velocity at release.  We show an example in \Figno\bartime,  where
the $k$-mode is chosen to be near a zero point
of the acoustic density oscillation at $\eta_d$.  
The rapid regeneration of density
fluctuations via velocity overshoot 
is due to the fact that zeros of the density oscillation
are maxima of the velocity oscillation.   The peaks in the matter power
spectrum due to baryonic acoustic oscillations therefore occur at
$$
k_{b j} r_s(\eta_d) = \cases {(j-1/2)\pi, & adiabatic \cr
			   j\pi, & isocurvature \cr }  
\eqn\eqnBaryonPeak
$$
and are roughly $\pi / 2$ out of phase with the corresponding CMB fluctuations.

To obtain the amplitude of the acoustic fluctuations, we must also
consider damping effects.  Photon diffusion in the tight coupling
regime damps baryon fluctuations as well due to Compton drag,
\ie\ via the Silk mechanism (Silk 1968).   
Analogous to the photon case,  we can construct the
drag visibility function $\Vb$ out of the drag optical depth $\tau_d$.
The acoustic visibility function then becomes $\hat \Vb = \Vb 
\exp[-(k/k_D)^2]$ (see \Figno\dampvis).  
Similarly, the net damping as a function of scale
is described by 
$$
\Db(k) = \int_0^{\eta_0} d\eta \hat \Vb(\eta,k) 
 = \int_0^{\eta_0} d\eta \Vb(\eta) e^{-[k/k_D(\eta)]^2} 
  \approx e^{-(k/k_S)^{m_S}}, 
\eqn\eqnSilk
$$
where the approximation is valid through the first decade of damping.
Analytic fitting formula for $k_S$ and $m_S$ are given in Appendix E 
eqns.~(E-9) (E-10).
As is the case with the photons, the latter accounts for the
width of the visibility function and is almost independent of
cosmological parameters.  Note that the Silk damping length is not
the same as the photon damping length despite the underlying
similarity in cause.  The small difference between last scattering 
and the drag epoch can alter it significantly
due 
to the rapid change in $k_D$ around recombination (see \Figno\dampvis).
Moreover, the two scale differently with the baryon and matter content.

Together with the horizon crossing boost from Appendix B, this defines the 
contribution to the matter transfer function of the acoustic oscillations
(see \Figno\trans). Explicit expressions for the three scenarios shown
are given in Appendix D eqns.~(D-22) (D-23).  
In the isocurvature BDM models, the transfer
function at small scales is dominated by the initial entropy 
fluctuation $S(0,k)$.  Furthermore, unlike the photon oscillations, 
baryon 
oscillations may survive early reionization if it occurs
more than an expansion time after the drag epoch.  In this case, the baryonic
oscillations are subsequently surrounded by a homogeneous and 
isotropic distribution of photons.
They then represent entropy perturbations and are not damped by 
further photon
diffusion.  Even in the rather unphysical event 
of near instantaneous reionization, baryonic oscillations may survive
if reionization is 
accompanied by the formation of a significant fraction of compact baryonic
objects (see \eg\ Gnedin \& Ostriker 1990).

If the model contains cold dark matter, baryons suffer an additional
effect.  After the drag epoch, they fall into potential wells established
by the CDM.  If the CDM to baryon ratio is high, this effect will dominate 
over the velocity overshoot of the acoustic oscillations.  To quantify
this effect, we need to consider the evolution of CDM fluctuations. 

\section{5. Cold Dark Matter Fluctuations}

\topinsert
\centerline{
\hskip -0.5truecm
\epsfxsize=3.0in \epsfbox{cdmtimea.epsf} \hskip 0.5 truecm
\epsfxsize=3.0in \epsfbox{cdmtimeb.epsf}
}

\baselineskip=12truept \leftskip=3truepc \rightskip=3truepc 
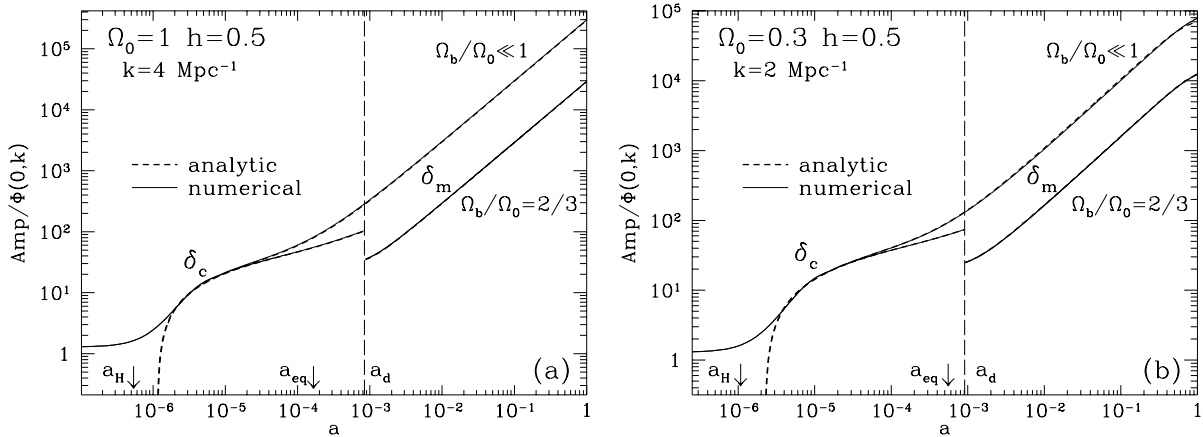
\noindent {\bf \Figno\cdmtime.} CDM and matter fluctuation
time evolution.  The cold dark
matter fluctuations are constant outside the horizon scale and experience
a boost into a logarithmically growing mode at horizon 
crossing in the radiation dominated
epoch.  
Between equality
and the drag epoch, the presence of baryons suppress the growth rate 
of CDM fluctuations [see eqn.~(7)]. 
By lowering $\Omega_0 h^2$, this region of suppression can 
be reduced for fixed $\Omega_b/\Omega_0$, {\it cf.} (a) and (b).
After the drag epoch, we plot $\delta_m = (\delta_b\rho_b +
\delta_c\rho_c)/(\rho_b+\rho_c)$ rather than $\delta_c$ since
the two components thereafter contribute similarly to the
total growth.  
Baryons also lower the contribution of $\delta_c$ to $\delta_m$ at the
drag epoch.
\endinsert
Let us begin with the evolution of CDM fluctuations before the
drag epoch.
As shown in \S 2 and Appendix B2, CDM fluctuations are given a 
kick at horizon crossing that sends them into a logarithmicly growing mode.  
As the universe becomes matter dominated, this stimulates power-law growing
and decaying modes $\delta_c \propto a^p$. 
If CDM dominates the non-relativistic matter, $p=\{ 1, -3/2 \}$.  However
if the baryon fraction is significant, the power law is modified. 
To first order in $\Omega_b/\Omega_0$, 
$$
p = \left\{ 1 - {3 \over 5}{\Omega_b \over \Omega_0} , \quad  -{3 \over 2}
\left[ 1 - {2 \over 5} {\Omega_b \over \Omega_0} \right] \right\}.
\eqn\eqnPowerApprox
$$ 
CDM growth is thus
inhibited by the presence of baryons.  We give the exact solution to
the evolution equation from horizon crossing to $z_d$ in Appendix D.
Because of the complexity of these expressions, it is also useful 
and instructive
to obtain approximate scaling relations.
If $a_d \gg a_{eq}$ and $\Omega_b /\Omega_0 \ll 1$, the main
effect is an
amplitude reduction of the CDM density perturbation 
$\delta_c(\eta_d,k)$ by approximately
$(a_d / a_{eq})^{-0.6 \Omega_b/\Omega_0}
\approx (24\Omega_0 h^2)^{-0.6 \Omega_b/\Omega_0}$ (see also Fig.~10).  
As $\Omega_0 h^2$ is lowered, the drag epoch
recedes into the radiation domination epoch and the regime where the growth rate
is affected $a_{eq} < a < a_d$ vanishes. 
 
At the drag epoch, the baryons are released from the photons and behave 
dynamically as if they were CDM. 
Since baryonic infall into CDM wells subsequently contributes to the
self-gravity of the matter, the growth rates again become
$p = \{1, -3/2 \}$ regardless of the baryon fraction.  We thus
follow the total non-relativistic matter perturbation $\delta_m$
after the drag epoch.  The relative
contribution of the CDM fluctuations at the drag epoch to the total
non-relativistic matter fluctuations $\delta_m$ scales as 
$1-\Omega_b/\Omega_0$.  
A good fit to the net suppression is given by the form
$$
\delta_m \approx  \alpha
\left( 1 - {\Omega_b \over \Omega_0} \right) 
	\times 
	\lim_{\Omega_b \rightarrow 0} \delta_m,
\eqn\eqnSuppression
$$
where $\alpha \approx (47 \Omega_0 h^2)^{-0.67\Omega_b/\Omega_0}$
for low $\Omega_b/\Omega_0 \simlt 0.5$ and high $\Omega_0 h^2 \gg 0.03$. 
The change in the coefficients from the naive scaling relation 
is due to detailed matching of growing and decaying modes [see Appendix D
eqn.~(D-18)].
A highly accurate fitting formula for $\alpha$ in the general case
is given in Appendix E, eqn.~(E-12). 
For the $\Omega_b
\rightarrow 0$ limit, an  exact expression in terms of elementary
functions is given in Appendix D which improves the $10\%$ accuracy
of the standard BBKS fitting formula to better than 
the $1\%$ level at small
scales.  

In \Figno\cdmtime, we show the resulting
evolution of a scale under the Silk damping length for which the
baryon fluctuations at the drag epoch are negligible.  
To extend the scaling of equation \eqnSuppression\ to larger scales
for \Figno\trans,
we have employed a generalization of the 
BBKS fitting function given in Appendix D.
Notice that
\smallskip
\refs (1) the change in the growth rate between equality and the drag
epoch, and

\refs (2) the fractional contribution of $\delta_c$ to $\delta_m$
at the drag epoch,

\smallskip
\noindent both play 
a significant 
role in suppressing the final amplitude of matter fluctuations.
Combined with the acoustic contributions from \S 4, this completes
the matter transfer function in CDM models.

We can now address the question of when acoustic oscillations 
are prominent
in the matter transfer function. The main effect is 
simply due to the density ratio $\rho_b/\rho_c = \Omega_b/\Omega_c$.
However, the acoustic and cold dark
matter contributions have a different dependence on scale.  Relative
to the cold dark matter, acoustic contributions scale as $(k\eta_d) \Db$
due to the velocity overshoot and diffusion damping factors.
Since $\Db$ encorporates an exponential cut off at the Silk 
scale $k_S$ and velocity overshoot weights the spectrum 
toward small scales,
acoustic contributions will be most visible just above the Silk scale.
The relative contribution to the matter transfer function will therefore
scale as $k_S \eta_d$.  Acoustic contributions increase in 
prominence if the Silk scale is small compared with the horizon 
at the drag epoch, \ie\ in the high $\Omega_b h^2$ case.  By including
the suppression factor from equation \eqnSuppression\ and numerical factors
from Appendix B and D, the maximum ratio of the acoustic amplitude to the
CDM contribution 
in the transfer function scales crudely as [see Appendix D, eqn.~(D-31)]
$$
0.4 k_S {\Omega_b \over \Omega_c}
	(\Omega_0 h^2)^{-1}
	(1+24\Omega_0 h^2)^{-1/2} (1+32\Omega_b h^2)^{-3/4} \alpha^{-1},
\eqn\eqnTransRatio
$$
where $k_S$, here in Mpc$^{-1}$, 
and $\alpha$ are given explicitly in Appendix E.  
This relation encorporates:
\smallskip

\refs (1) Baryon acoustic oscillations;

\refs (2) Baryon decoupling;

\refs (3) Silk damping;

\refs (4) Velocity overshoot;

\refs (5) Baryon gravitational instability;

\refs (6) Baryon infall into CDM wells whose depth depends on

\refsq (a) Acoustic feedback into the potential,

\refsq (b) CDM self-gravity vs. the expansion rate,

\noindent and combines them in a consistent manner.

\smallskip

\section{6. Discussion} 

We have established a framework for treating small scale fluctuations
in the realistic case of a coupled multifluid system.
In the limit that fluctuations crossed
the horizon in the radiation dominated epoch, closed-form analytic
solutions are available.  The fundamental elements uncovered by this 
approach are the boost at horizon crossing due 
to infall and dilation effects from
potential decay, the source-free solution of the component evolution equations,
the baryon drag on the photons, and the Compton drag on the baryons.  
Together they establish general consistency relations between the matter
and the radiation power spectra as well as expose their sensitivity
to changes in the background model. 

Unlike the case of the matter fluctuations, decay in the potential results in an 
amplification of acoustic oscillations in the CMB.  
These opposing manifestations of the same physical effect provide
both a measure and a powerful consistency check on the redshift of
equality.
Baryon drag displaces the zero point of the acoustic oscillations which 
remains as a temperature shift even after the oscillations have damped by
photon diffusion.  Last scattering marks the end of the baryon drag
epoch at which the acoustic oscillations with their characteristic
diffusion damping scale are frozen into the photon spectrum.  Correspondingly,
at the end of the Compton drag epoch, the Silk-damped baryonic acoustic
oscillations are frozen into the matter spectrum.  We have provided 
convenient analytic
fitting formulae for these quantities as a function of the matter and
baryon content. These may be useful for the extraction of cosmological
information from the CMB and matter 
power spectrum once observations become available.  

The photon-baryon system also affects the growth of CDM fluctuations.
It first stimulates growth through the horizon crossing boost.  
Subsequently, it affects the balance of the 
growth inhibiting expansion to the self-gravity of the CDM. 
We have obtained an analytic solution for adiabatic initial conditions 
and the exact general solution for the two effects respectively.
To simplify these expressions, we have also provided an accurate fitting
form for the transfer function in terms of elementary functions.
Growth suppression due to the presence of baryons has implications for
the first generation of structure.  The expressions derived here 
remain valid for linear perturbations to an arbitrarily small scale
where direct numerical calculations are impractical.

Baryonic acoustic oscillations are of course not prominent in presently
popular models where $\Omega_b/\Omega_0 \ll 1$ due to the extra growth
of the CDM fluctuations between horizon crossing and the drag epoch.  
However, they serve as
a useful complement to their CMB counterpart if either 
$\Omega_0 h^2\simlt 0.05$ 
or big bang nucleosynthesis constraints are too stringent (Gnedin
\& Ostriker 1990).  
Indeed there are tentative indications from cluster inventories that
the baryon fraction may be as high as 15\%  (White \etal\ 1993).  
The presence or absence of acoustic oscillations
in the observations of the CMB and large scale structure will in the
future provide a robust distinction between general classes of
scenarios:

\refs (1) Oscillations in CMB and matter power spectra: 
standard recombination with $\Omega_b \simgt \Omega_c$;

\refs (2) Oscillations in CMB alone: standard recombination with
$\Omega_b \ll \Omega_c$;

\refs (3) Oscillations in matter power spectra alone: early reionization
with $\Omega_b \simgt \Omega_c$.

\noindent
Large scale structure observations already suggest there are no
dramatic oscillations in the matter power spectrum as would be 
the case for (1) and (3) if $\Omega_b \gg \Omega_c$  (Peacock
\& Dodds 1994). 
However low amplitude oscillations, as might be expected if $\Omega_b 
\sim \Omega_c$, remain possible.  Of course, the exact form that these
oscillations will take in the observations 
depends on issues such as redshift space
distortions and non-linear corrections.  

If oscillations are discovered in {\it neither} spectra, the most
natural conclusion is our universe has $\Omega_b \ll \Omega_c$ and
suffered early reionization.  However other possibilities include
the formation of perturbations after recombination, reionization
within an expansion time after the drag epoch, equal or random stimulation of
adiabatic and isocurvature
mode acoustic fluctuations.\footnote{$^{\rm\dag}$}{This may be accomplished
by balancing the initial conditions or hypothesizing a gravitational
forcing potential that is
external to the linear photon-baryon-neutrino-CDM system, {\it e.g.}
cosmological defects.}
These scenarios can be distinguished by
measuring the location of the damping scale.  All scenarios obeying
standard recombination, 
regardless of the presence of actual peak-like structures, follows
the scalings for the damping length 
 discussed here.  In all reionized models, the horizon at last scattering 
and at the drag epoch marks the damping scale for the photons
and baryons respectively.  

If acoustic oscillations are discovered in {\it both} the CMB and large
scale structure power spectra, we will possess a strong consistency test
for the dynamics of the expansion, \ie\ a combination of 
the matter content, curvature, and cosmological
constant, as well as the adiabatic or isocurvature 
nature of the initial fluctuations.
Furthermore, the two contain complementary information on 
several specific fundamental
cosmological parameters. The scale of the peaks 
in the matter power spectrum are mainly determined by the matter content
$\Omega_0 h^2$ whereas the angular scale of the 
 CMB peaks is mostly sensitive to the curvature
$1-\Omega_0 -\Omega_\Lambda$.  The two damping lengths also probe 
different combinations of $\Omega_0 h^2$ and $\Omega_b h^2$.  
The ratio of the peak heights to the underlying CDM contribution
in the matter power spectrum
probes $\Omega_b/\Omega_0$. 
Furthermore, by comparing similar scales, dependence on the initial power 
spectrum can be eliminated providing a clean test of the whole gravitational
instability paradigm.  

\section{Acknowledgments} 

We would like to acknowledge useful discussions with J.R. Bond, 
U. Seljak, J. Silk, \& P. Steinhardt.  W.H. would like to thank
J.R. Bond for calling to his attention the enhancement of
diffusion damping through polarization and M. White for pointing out 
several typos in the draft.  W.H. was supported by grants 
from the NSF and W.M. Keck Foundation.

\eject 
\section{Appendix A: Tight Coupling Limit}

\subsection{A.1. Evolution Equations}

The Fourier transform
of the Newtonian temperature fluctuation can be broken up into
Legendre moments, related to the direction cosines of the
photon momenta $\gamma_i$, $\Theta(\eta,\bk,\bg)
= \sum_\ell (-i)^\ell \Theta_\ell P_\ell(\bk \cdot \bg)$.  The evolution
equation for these moments is given by the Boltzmann hierarchy 
(Bond \& Efstathiou 1984, HSb)
$$
\eqalign{
\dot \Theta_0 & = -{k \over 3} \Theta_1 - \dot \Phi, \cr
\dot \Theta_1 & = k \left[\Theta_0 + \Psi - {2 \over 5} \Theta_2 \right]
		 - \dot \tau (\Theta_1 - V_b), \cr
\dot \Theta_2 & = k \left[{2 \over 3} \Theta_1 - {3 \over 7} \Theta_3 \right]
		 - \dot \tau \left( {9 \over 10} \Theta_2 - {1 \over 10}
		\Theta_2^Q - {1 \over 2} \Theta_0^Q \right), \cr
\dot \Theta_\ell & = k \left[ {\ell \over 2\ell -1} \Theta_{\ell - 1}
		      - {\ell +1 \over 2\ell + 3} \Theta_{\ell + 1} \right]
		      - \dot\tau \Theta_\ell,\quad {(\ell > 2)}
} 
\Aeqn\AeqnHierarchy
$$ 
if the ratio of the wavelength to the curvature scale is much smaller
than the angle considered, \ie\ 
$k \gg \ell\sqrt{-K}$
where $K= -H_0^2 (1-\Omega_0 - \Omega_\Lambda)$.  Recall that overdots
are derivatives with respect to conformal time $\eta = \int dt/a$.
Here $\Theta^Q_\ell$ is the CMB temperature perturbation in the Stokes
parameter $Q$.  It accounts for polarization generated by Compton scattering
of anisotropic radiation. 
The metric fluctuations in the mode are given by $g_{00} = -(1 + 2\Psi Y)$ and
$g_{ij} = a^2(1+2 \Phi Y) \gamma_{ij}$,  
where $\gamma_{ij}$ is the three
metric on a surface of constant curvature and $Y$ is a plane wave 
$e^{i {\bf k} \cdot {\bf x}}$
in flat space or more generally a $k$-eigenfunction of the Laplacian.  
The presence of the 
curvature perturbation $\Phi$ in the monopole equation represents the dilation
effect.  The form of the metric shows that it has the same origin as
the photon redshift with the expansion.   The gravitational potential
$\Psi$ in the dipole or velocity equation accounts for gravitational infall
or redshift.  

The tight coupling approximation
assumes that the Compton scattering rate
$\dot \tau$ is sufficiently rapid to equilibrate changes in the
photon-baryon fluid.
It is an expansion in the 
Compton scattering time $\dot\tau^{-1}$, or more specifically the inverse of
the optical depth through a wavelength $\dot \tau/k$ and through a period
of the oscillation $\dot \tau / \omega = \dot \tau / kc_s > \dot\tau/k$, 
where
$$
c_s = {1 \over \sqrt{3(1 +R)}}
\Aeqn\AeqnSound
$$
is the photon-baryon sound speed with 
$R = {3 \over 4} \rho_b/ \rho_\gamma$.
To first
order, only the $\ell = 0$ monopole (with density fluctuation
 $\delta_\gamma = 4\Theta_0$) 
and $\ell = 1$ dipole (with bulk velocity $V_\gamma = \Theta_1$) survive
and one obtains the forced oscillator equation for acoustic waves in
the photon-baryon fluid 
(HSa).\footnote{$^{\rm\dag}$}{In HSa and HSb, we employed
a hybrid gauge or ``gauge invariant'' representation of density
fluctuations for computational convenience.  Since there are no benefits
of this choice below the horizon, we
work entirely in the Newtonian gauge in this paper.  Only the definition
of density fluctuations is affected: 
the total matter gauge 
$\Delta_X  = \delta_X + 3 {\dot a \over a} (1 + p_X/\rho_X) V_T/k$ where 
$X$ represents any of the particle components.}
To second order, the acoustic oscillations of the monopole and dipole 
are damped due to the imperfect coupling between the photons and baryons.
Photon diffusion creates 
heat conduction through $\Theta_1 - V_b$ and shear viscosity through
$\Theta_2$ (Weinberg 1972, Bond 1995).

To close these equations, we need the continuity and Euler equations for the 
baryons
$$
\eqalign{
\dot \delta_b & = - kV_b - 3\dot \Phi, \cr
\dot V_b & = -{\dot a \over a} V_b + k\Psi + \dot \tau (\Theta_1 
	- V_b)/R, \cr
}
\Aeqn\AeqnBaryon
$$
the polarization hierarchy equations for the CMB 
(Bond \& Efstathiou 1984, Kosowsky 1995)
$$
\eqalign{
\dot \Theta_0^Q & = -{k \over 3} \Theta_1^Q - \dot \tau 
		    \left[ {1 \over 2} \Theta_0^Q - {1 \over 10}
		    (\Theta_2 + \Theta_2^Q) \right], \cr
\dot \Theta_1^Q & = k \left[\Theta_0^Q - {2 \over 5} \Theta_2^Q \right]
		 - \dot \tau \Theta_1, \cr
\dot \Theta_2^Q & = k \left[{2 \over 3} \Theta_1^Q 
		 - {3 \over 7} \Theta_3^Q \right]
		 - \dot \tau \left( {9 \over 10} \Theta_2^Q - {1 \over 10}
		\Theta_2 - {1 \over 2} \Theta_0^Q \right), \cr
\dot \Theta_\ell^Q & = k \left[ {\ell \over 2\ell -1} \Theta_{\ell - 1}^Q
		      - {\ell +1 \over 2\ell + 3} \Theta_{\ell + 1}^Q \right]
		      - \dot\tau \Theta_\ell^Q, \quad {(\ell > 2)}
} 
\Aeqn\AeqnQHierarchy
$$ 
and the Einstein-Poisson equations for the metric or potential perturbations
$$
\eqalign{
k^2 \Phi & = 4\pi G a^2 \rho_T [ \delta_T + 
	3{ \dot a \over a} (1+w_T) {V_T \over k}], \cr
k^2(\Phi + \Psi) &  = - 8\pi G a^2 p_T \Pi_T,
}
\Aeqn\AeqnPoisson
$$
if $k \gg \sqrt{-K}$.
Here $w_T = p_T / \rho_T$ where the subscript $T$ denotes
the total matter including all particle species 
and the total anisotropic stress is related to the radiation 
quadrupoles as   
$$
\Pi_T = {12 \over 5} [\Theta_2 + N_2],
\Aeqn\AeqnPi
$$
with $N_2$ as the neutrino temperature quadrupole.   Notice that
the baryon continuity equation can be rewritten as $\dot \delta_b 
= -k(V_b-\Theta_1) + 3\dot\Theta_0$ since dilation effects on the
photon temperature and baryon density fluctuations are analogous.  
This represent adiabatic evolution if $V_b = \Theta_1$.

\topinsert
\centerline{
\epsfxsize=3.5in \epsfbox{f2.epsf}
}

\baselineskip=12truept \leftskip=3truepc \rightskip=3truepc 
\noindent {\bf \Figno\ftwo.} 
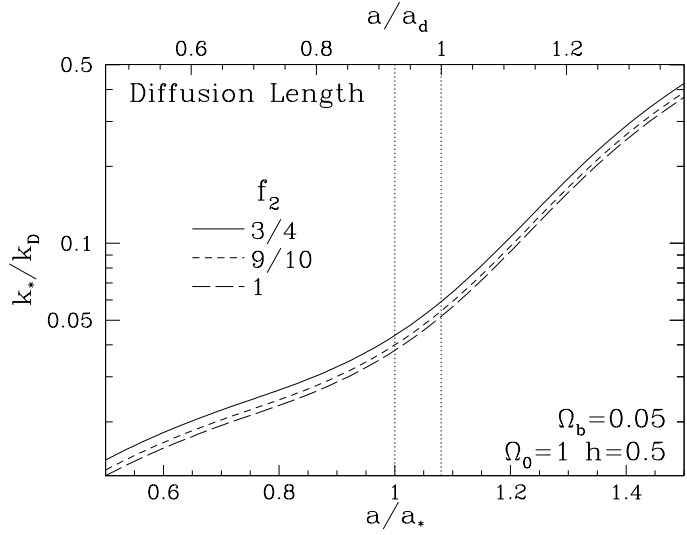
Photon diffusion scale.  The photon diffusion scale grows rapidly
near last scattering due to the increasing mean free path of the
photons but remains well under the horizon scale
$k_*^{-1} = (\dot a/a)|_{a_*}$ at last scattering.   The small
difference between $a_*$ and $a_d$ is sufficient to cause a 
significant difference in  the effective damping if $\Omega_b h^2$ differs
substantially from the crossover point $0.03$.  The inclusion of
the angular dependence of Compton scattering enhances damping by
a small factor $f_2 = 9/10$ as does the further inclusion of polarization
$f_2 = 3/4$.
\endinsert
\subsection{A.2. Acoustic Dispersion Relation}

Let us derive the dispersion relation for acoustic oscillations in the tight
coupling limit.  Consider first the effects of polarization.
Since only the polarization monopole $\Theta^Q_0$ and quadrupole
$\Theta^Q_2$ feed back into
the temperature fluctuations, we may immediately expand the polarization
hierarchy in $\dot \tau^{-1}$ to obtain
$$
\Theta_2^Q = \Theta_0^Q = {1 \over 4} \Theta_2,
\Aeqn\AeqnPiQ
$$
which simplifies the temperature quadrupole evolution of \Adis\AeqnHierarchy\ to
$$
\dot \Theta_2  = k \left[{2 \over 3} \Theta_1 - {3 \over 7} \Theta_3 \right]
		 - \dot \tau f_2 \Theta_2, 
\Aeqn\AeqnQuadrupole
$$
where $f_2 = {3 \over 4}$.  Other approximations commonly used are 
$f_2 = {9 \over 10}$ for unpolarized radiation 
(Chibisov 1972)
and $f_2 = 1$ for further neglecting the angular dependence of Compton 
scattering  (Weinberg 1972, Peebles 1980, HSa).
We keep the factor $f_2$ implicit so that the separation
of effects can be read directly off the final results.  Expanding
the quadrupole equation \Adis\AeqnQuadrupole\ to first order in $\dot\tau^{-1}$ 
we obtain 
$$ 
\Theta_2 = \dot\tau^{-1} f_2^{-1} {2 \over 3} k\Theta_1.
\Aeqn\AeqnQuad 
$$
A second-order expansion for the quadrupole is not necessary since its 
effect on the fluid equations through $\dot \Theta_1$  
is already of first order in $\dot \tau^{-1}$.

On the other hand, a {\it second} order expansion 
of the baryon Euler equation
is necessary.  Let us try a solution of the form $\Theta_1 \propto
\exp i\int \omega d\eta$ and ignore variations on the expansion time scale
$\dot a / a$ in comparison with those at the oscillation frequency $\omega$.  
The electron velocity, obtained by iteration of the Euler equation, is to
second order
$$
V_b = \Theta_1 - \dot\tau^{-1} R[i\omega\Theta_1
	 - k\Psi] - \dot\tau^{-2} R^2 \omega^2 \Theta_1. 
\Aeqn\AeqnVb
$$
Substituting this into the photon dipole equation \Adis\AeqnHierarchy\ and
eliminating the zeroth order term yields
$$ 
i\omega (1+R) \Theta_1 = k[\Theta_0 + (1+R)\Psi] - \dot\tau^{-1} R^2 
	\omega^2 \Theta_1 - {4 \over 15} \dot\tau^{-1} f_2^{-1} k^2 \Theta_1.
\Aeqn\AeqnOmegaa
$$
This suggests that we try a solution of the form $\Theta_0 + (1+R)\Psi
\propto \exp i\int\omega d\eta$.  Employing the monopole equation of
\Adis\AeqnHierarchy\ and again assuming that variations at the oscillation 
frequency are sufficiently rapid that changes in $R$, $\Phi$, and $\Psi$ 
can be neglected, we obtain
$$
(1+R)\omega^2 = {k^2 \over 3} + i \dot\tau^{-1} \omega \left( R^2 \omega^2
	         + {4 \over 15} k^2 f_2^{-1} \right).
\Aeqn\AeqnOmegab
$$
With the first order relation $\omega^2 = k^2/3(1+R)$, the solution
to the resultant quadratic equation is
$$
\omega = \pm {k \over \sqrt{3(1+R)}} + i {1 \over 6} k^2 \dot\tau^{-1} 
	\left[ {R^2 \over (1+R)^2} + {4 \over 5} f_2^{-1} {1 \over 1+R}
	\right]. 
\Aeqn\AeqnDispersion
$$
Thus to second order acoustic oscillations are damped as 
$\exp [ -(k/k_D)^2]$ with the damping length 
(Weinberg 1972, Kaiser 1983, Bond 1995)
$$ 
k_D^{-2} = {1 \over 6} \int d\eta {1 \over \dot\tau} 
{R^2 + 4f_2^{-1} (1+R)/5 \over (1+R)^2}.
\Aeqn\AeqnDamp
$$
If $R \simlt 1$, 
polarization and the angular dependence of Compton scattering
enhances damping through the generation of viscosity
(see \Figno\ftwo). 
Viscosity is related to photon diffusion because the
quadrupole and higher moments are generated as photons from regions of
different temperatures meet.

\topinsert
\centerline{
\hskip -0.5truecm
\epsfxsize=3.0in \epsfbox{rpsia.epsf} \hskip 0.5truecm
\epsfxsize=3.0in \epsfbox{rpsib.epsf}
}

\baselineskip=12truept \leftskip=3truepc \rightskip=3truepc 
\noindent {\bf \Figno\rpsi.}
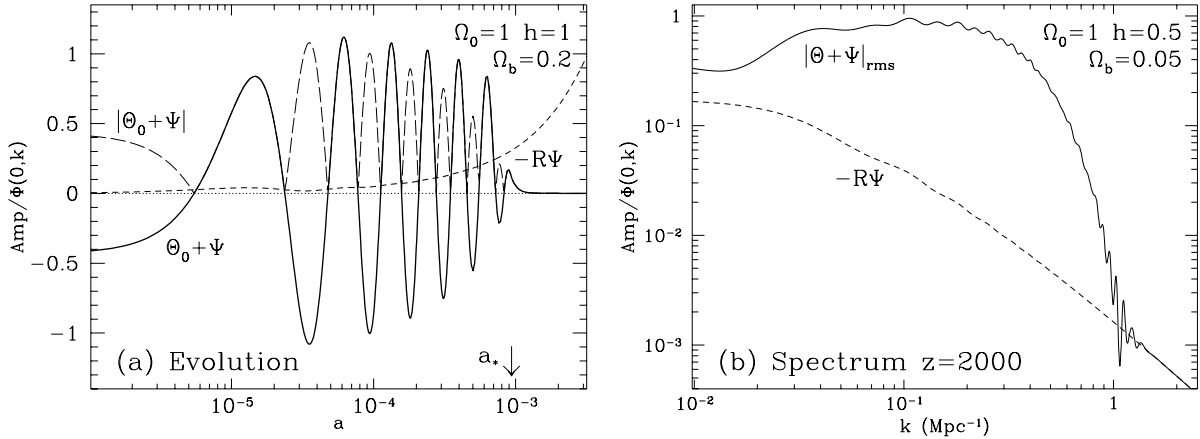
Baryon drag effect in adiabatic CDM models. 
(a) Baryons cause a drag effect on the photons 
leading to a temperature enhancement of $-R\Psi = |R\Psi|$ inside
potential wells which shifts the zero point of the oscillation 
(short dashed lines).  
(b) This contribution yields alternating peak heights in the rms 
and is also retained after 
diffusion damping.  Here numerical results are displayed.  
\endinsert
\subsection{A.3. Baryon Drag and the Adiabatic Invariant}

The temperature perturbation oscillates around $\Theta_0 + (1+R)\Psi = 0$
due to the baryon drag effect.  
This can be more easily understood by examining the first order equation 
from which it originates,
$$
(1+R) \ddot \Theta_0 + {k^2 \over 3} \Theta_0 \approx -{k^2 \over 3}(1+R)\Psi,
\Aeqn\AeqnAcous
$$
ignoring slow changes in $R$, $\Phi$ and $\Psi$ 
from the expansion.  Notice that $m_{\rm eff} = (1+R)$ plays the role of the 
effective mass of the oscillator and the gravitational potential provides the
effective acceleration through infall.  The photon pressure acts as 
the restoring force and is independent of the baryon content $R$.  

Equation \Adis\AeqnAcous\ has the immediate
solution 
$$
\Theta_0 = [\Theta_0(0,k) + (1+R)\Psi] \cos(\omega\eta) + {1 \over \omega}
\dot\Theta_0(0,k) \sin(\omega\eta) - (1+R)\Psi,
\Aeqn\AeqnSimple
$$
where the frequency 
$\omega = k/\sqrt{3 m_{\rm eff}} = k/\sqrt{3(1+R)}$ in agreement
with the first order dispersion relation of \Adis\AeqnDispersion.  
This solution describes an oscillator
whose zero point has been displaced by $-m_{\rm eff} \Psi = -(1+R)\Psi$
due to the gravitational force.  The
photons, although massless, suffer infall effects due to gravitational 
blueshift.  Since this is exactly cancelled as the photons
stream out of the potential wells, we can consider $\Theta_0+\Psi$ as
the effective temperature perturbation.   Thus this part of the zero point
shift has no net effect.  However, the baryons also contribute to the effective
mass of the fluid.  Since the photons and baryons are tightly coupled, 
baryonic infall drags the photons into potential wells and contributes
$-R\Psi$ to the displacement.  Notice that baryon drag also
increases the amplitude of the cosine oscillation since the initial conditions
$\Theta(0,k)$ represent a greater displacement from the zero point.  
Thus baryon drag accounts for both the alternating  
peak heights of the acoustic oscillations and their enhancement
with $\Omega_b h^2$ (HSa).  
After diffusion damping has eliminated the oscillations
themselves, the zero point shift $-R\Psi$ remains.  Of course,
for adiabatic BDM models $\Psi$ is also reduced to zero by the
diffusion.

In reality, the variation of oscillator parameters, such as the
effective mass and gravitational force, on the expansion time
scale cannot be ignored over many periods of the oscillation.
From equations \Adis\AeqnHierarchy\ and \Adis\AeqnBaryon, 
the full first order equation is
$$
{d \over d\eta} (1+R) \dot\Theta_0 + {k^2 \over 3}\Theta_0 = -{k^2 \over 3}
(1+R)\Psi - {d \over d\eta} (1+R)\dot\Phi,
\Aeqn\AeqnFirstOrder
$$
where the addition of the space curvature term comes from the dilation effect
$\dot \Theta_0 = - \dot \Phi$.  Notice that the left hand side is
precisely the equation of an oscillator with a time-varying mass
$m_{\rm eff} = (1+R)$.  The homogeneous equation can be solved 
by employing the fact that variations over a single
period
of the oscillation are small. The adiabatic invariant associated with an
oscillator is given by the energy $E = {1 \over 2} m_{\rm eff} \omega^2
A^2$ over the frequency $\omega$.  The amplitude therefore scales as 
$A \propto \omega^{1/2} \propto (1+R)^{-1/4}$.  This yields fundamental
solutions of the form $(1+R)^{-1/4} \exp(\pm i\int \omega d\eta)$
(Peebles \& Yu 1970). 
The phase integral can be performed analytically 
$$
\int \omega d\eta = k\int c_s d\eta = k r_s,
$$ 
where $r_s$ is the
sound horizon given in equation \eqnSound.

\subsection{A.4. Summary}

\noindent
Several points are worth emphasising:  
\smallskip
\refs
(1) The first order dispersion relation
for acoustic oscillations is $\omega = kc_s = k/\sqrt{3(1+R)}$.

\refs
(2) Slow changes in the baryonic contribution to the effective mass cause
the temperature oscillation to decay as $(1+R)^{-1/4}$. 

\refs
(3) The oscillation phase is related to the sound horizon $r_s = \int 
c_s d \eta$ by $\int \omega d\eta = kr_s$.  

\refs
(4) Photon diffusion alters the dispersion relation and leads to
exponential damping.

\refs
(5) The damping length increases roughly as $\lambda_D \sim k_D^{-1} \sim
\sqrt{\eta / \dot \tau}$ or the geometric mean of the conformal time and
the Compton mean free path as one expects of a random walk. 

\refs
(6) If it is well under the horizon, 
the damping length surpasses the wavelength when $\dot \tau/k = k\eta \gg 1$
or when the optical depth through a wavelength is still high. 

\refs
(7) The angular dependence of Compton scattering and polarization increases
the damping length when $R \simlt 1$.

\refs
(8) The zero point of oscillations in $\Theta_0 + \Psi$ 
is $-R\Psi$ due to baryon drag and remains as a temperature shift 
after diffusion damping.  
\smallskip

\noindent

\section{Appendix B. Horizon Crossing} 

\subsection{B.1. Photon-Baryon System}

The amplitude of the acoustic oscillation is determined by the growth of
fluctuations before and in particular {\it during} the epoch when the 
scale crosses the horizon. 
Since the second-order tight-coupling expansion
just yields a multiplicative diffusion damping factor, let us obtain the
first order solution which we denote by overhats.  
The full solution can be constructed through the relations
$$
\eqalign{
\Theta_0 + \Psi &= [\hat \Theta_0 + (1+R)\Psi] \exp[-(k/k_D)^2] - R\Psi, \cr
\Theta_1 & = \hat \Theta_1 \exp[-(k/k_D)^2], \cr
\delta_b & = 3\Theta_0 + S(0,k), \cr
kV_b & = k\Theta_1 = -\dot\Theta_0 - \dot\Phi,   }
\Beqn\BeqnHat
$$
where recall $S(0,k)$ is the initial entropy
perturbation in the photon-baryon system $S(0,k) = \delta_b(0,k) - 
{3 \over 4} \delta_\gamma(0,k)$.
Near or above the horizon at $\eta_*$, 
where the sources
from potential growth and decay drive the oscillator, the full formalism
of HSa is needed to follow the fluctuations accurately.  
However, we are only interested in small scale fluctuations which
enter the horizon well before last scattering.
Since radiation pressure prevents the growth of fluctuations in the
radiation dominated epoch, the 
gravitational potentials decay away after horizon crossing.  
On small scales, the acoustic
oscillations therefore experience a boost at horizon crossing and thereafter
settle into pure modes $
\Theta_0(\eta,k) = (1+R)^{-1/4} [ C_A(k) \cos(kr_s) + C_I(k) \sin(kr_s) ] $.
Equation \Adis\AeqnFirstOrder\ 
yields exact solutions for $C_A$ and $C_I$ for the fundamental
adiabatic and isocurvature modes of the fluctuation if anisotropic stress
$\Pi_T$ is
ignored so that $\Psi = -\Phi$.  The adiabatic mode arises from an initial
curvature perturbation $\Phi(0,k)$, whereas the isocurvature mode from
an initial entropy perturbation $S(0,k)$.  The solutions are
(Kodama \& Sasaki 1986, HSb)
$$
\eqalign{
\lim_{\Pi \rightarrow 0} C_A(k) &= {3 \over 2}\Phi(0,k), 
	\qquad {\rm adiabatic} \cr
\lim_{\Pi \rightarrow 0} C_I(k) & = -{\sqrt{6} \over 4} {k_{eq} \over k} S(0,k), 
	\qquad {\rm isocurvature} 
}
\Beqn\BeqnPiZero
$$
where $k_{eq} = (2 \Omega_0 H_0^2 /a_{eq})^{1/2} = 9.67 \times 10^{-2} 
\Omega_0 h^2 (1 - f_\nu)^{1/2} \Theta_{2.7}^{-2}$ Mpc$^{-1}$ is
the wavenumber that passes the horizon at equality.  
Recall that $f_\nu = \rho_\nu/(\rho_\gamma + \rho_\nu)$.
The two modes stimulate pure cosine and sine modes since the gravitational
forcing function yields near resonant driving with the 
phase fixed by $\Phi(0,k) = $ constant and $\Phi(0,k)=0$ respectively
(HSb, Hu \& White 1996).  The amplitude of the
fluctuations is easy to understand qualitatively.  For the
adiabatic case,
$\Theta_0(0,k) = {1 \over 2} \Phi(0,k)$.  If photon streaming is ignored 
$\dot \Theta = -\dot \Phi$, the dilation effect would raise the amplitude
to $\Theta_0(\eta,k) = {1 \over 2} \Phi(0,k) - \Phi(\eta,k) + \Phi(0,k) 
= {3 \over 2}\Phi(0,k)$.  Note that a decaying potential {\it boosts}      
the acoustic amplitude due to the gravitational forcing effect. 
A similar analysis for the isocurvature mode accounting for potential
growth outside the horizon explains the isocurvature amplitude (HSb).

The effect of anisotropic stress can be considered as a perturbation
(HSa).  The dominant term comes from the neutrino quadrupole since
photon anisotropies are damped exponentially with optical depth before
last scattering.  The order of magnitude can be simply read off the 
initial conditions for the growing mode of the perturbation,  
$$
\Theta_0(0,k) = -{1 \over 2}\Psi(0,k) = {1 \over 2}
	\left( 1+ {2\over 5} f_\nu \right)^{-1} \Phi(0,k), \quad {\rm adiabatic}
\Beqn\BeqnICadi
$$
or $\Theta_0(0,k)=\Psi(0,k)=\Phi(0,k)=0$ and
$$
\eqalign{
\dot\Theta_0(0,k) & = {\sqrt{2} \over 16} k_{eq} 
\left( 1 + {2 \over 15} f_\nu \right)
S(0,k) ,  \cr
\dot\Phi(0,k) & ={\sqrt{2} \over 16} k_{eq}
\left( 1 +{2 \over 15} f_\nu 
	\right) S(0,k), \qquad {\rm isocurvature}\cr
\dot\Psi(0,k) & =-{\sqrt{2} \over 16} k_{eq}
\left( 1 - {6 \over 15} f_\nu  
	\right) 
	S(0,k). \cr
}
\Beqn\BeqnICiso
$$
Indeed, we find that the adiabatic amplitude is well approximated by
$$
C_A(k) = {3 \over 2}\left(1 + {2 \over 5} f_\nu \right)^{-1} \Phi(0,k), 
 \qquad {\rm adiabatic }
\Beqn\BeqnAmpAdiabatic
$$
and likewise
$$
C_I(k) = 
		-{\sqrt{6} \over 4} {k_{eq} \over k} 
		\left( 1 - {4 \over 15} f_\nu  \right) 
		S(0,k), \qquad {\rm isocurvature}
\Beqn\BeqnAmpIsocurvature
$$
for the isocurvature mode.
Explicitly, the first order acoustic solution is
$$
\eqalign{
\hat \Theta_0 + \Psi \approx \hat \Theta_0 & = (1+R)^{-1/4} [ 
	C_A\cos(kr_s) + C_I\sin(kr_s) ] ,
			\cr
\hat \Theta_1 = \hat V_b  & = -{\sqrt{3}}(1+R)^{-3/4} [ C_A \sin(kr_s) + 
		C_I \cos(kr_s) ].  \cr
}
\Beqn\BeqnAcoustic
$$
In \Figno\cmbtime, we display an adiabatic example.  In this
case,
it is also useful to compare the acoustic amplitude to the Sachs-Wolfe
effect (Sachs \& Wolfe 1967)
 $|\Theta+\Psi|_{rms}(\eta_0,k) = {1 \over 3}\Psi(\eta_0,k)$.
With the relation (see HSa eq.~A18),
$$
\Psi(\eta_0,k) = -{9 \over 10} \left( 1 + {4 \over 15}f_\nu \right)^{-1}
\left( 1 + {2 \over 5}f_\nu \right)^{-1} \Phi(0,k),
\Beqn\BeqnLSpsi
$$
valid at large scales $k \ll k_{eq}$, the relative amplitude becomes
$$
|3 C_A(k)/\Psi(\eta_0,k)| = 5 \left( 1 + {4 \over 15}f_\nu \right)^{-1},
\Beqn\BeqnSWBoost
$$ 
and represents a significant boost.

\subsection{B.2. CDM Component}

The evolution of the cold dark matter fluctuations in the presence of
acoustic oscillations is also interesting and relevant for determining
the small scale behavior of the matter transfer function (see Appendix D).   
The cold
dark matter evolution equations are of the same form as the baryon 
continuity and Euler equations save for the absence of coupling
to the photons,
$$
\ddot \delta_c + {\dot a \over a} \dot \delta_c = -k^2 \Psi - 3\ddot \Phi.
\Beqn\BeqnCDM
$$
In the radiation dominated epoch, the metric terms on
the right hand side are dominated by perturbations in the radiation and
may be considered as external driving forces.  The homogeneous equation
has two fundamental solutions $\delta_c \propto \{ \ln a, 1 \}$.
The particular solution is constructed via Green's method
$$
\delta_c = C_1 \ln a + C_2 + \int_0^{\eta} [\ln a' - \ln a] {a' \over \dot a'} 
(k^2 \Psi + 3\ddot \Phi) d\eta'.
\Beqn\BeqnGreenCDM
$$ 
Adiabatic initial conditions require $C_1=0$ and $C_2 = 3\Theta_0(0,k)$.  
Thus $\delta_c$ remains constant outside the horizon and then gets 
a kick from infall and dilation that generates a logarithmic growing 
mode.  Since the behavior of the potentials is self-similar in $k$,
\ie\ they are constant outside the horizon and decay to zero as $a^{-2}$
inside of it, their effect on $\delta_c$ is the same for all $k$. 
Once the potentials have decayed to zero, $\delta_c$ settles into the
logarithmic growing mode as
$$
\delta_c \approx I_1 \Phi(0,k) \ln \left(  I_2 {a \over a_H} 
	\right), 
\qquad a_H \ll a \ll a_{eq}
\Beqn\BeqnCDMrd
$$
where horizon crossing occurs at
$$
\eqalign{
{a_H \over a_{eq}} & = {1 + \sqrt{1+ 8(k/k_{eq})^2} \over 4(k/k_{eq})^2} \cr
	& \approx {\sqrt{2} \over 2} {k_{eq} \over k}.  \qquad k \gg k_{eq}
}
\Beqn\BeqnaH
$$ 
By numerical calculation of the integrals in \Bdis\BeqnGreenCDM, we obtain
$$
\eqalign{
I_1 &= 9.11 ( 1+ 0.128 f_\nu + 0.029 f_\nu^2 ), \cr
I_2 &= 0.594 ( 1 - 0.631 f_\nu + 0.284 f_\nu^2 ), 
}
\Beqn\BeqnCDMIntegrals
$$
valid at the 1\% or better level for the full range $0 \le f_\nu \le 1$. 
As we shall see in Appendix D, this solution can be joined onto the growing
mode in the matter dominated epoch to describe the full time evolution of
the CDM fluctuations.

\section{Appendix C. Decoupling}

\subsection{C.1 Photon Decoupling}

The tight coupling approximation is strictly valid only well before 
decoupling.  However, the acoustic modes may be simply joined onto
the free-streaming solutions once diffusion damping near decoupling
has been taken into account. 
The full Boltzmann hierarchy has the formal solution (see HSa eq.~11)
$$
[\Theta + \Psi](\eta_0,k,\mu) 
= \int_0^{\eta_0} d\eta [(\Theta_0 + \Psi - i\mu V_b) \dot\tau - \dot\Phi
+\dot\Psi] e^{-\tau} e^{ik\mu(\eta-\eta_0)} ,
\Ceqn\CeqnBoltzformal
$$
where $k\mu = \bk \cdot \bg$ and curvature has been neglected.  
The terms in parenthesis contribute at last
scattering due to weighting by the visibility function $\Vg = \dot\tau
e^{-\tau}$ and the ISW metric terms play a role between last scattering
and the present.  This formal solution is made practical by replacing 
the sources $\Theta_0$ and $V_b$ with their acoustic solution 
at last scattering.

It may seem that employing the tight coupling solution through decoupling
would lead to erroneous results.  In particular, the damping
approximation should break down if the optical depth through a wavelength
drops below unity.  However, let us examine the tight coupling solution
more carefully.  
Equation \Bdis\BeqnHat\ implies
$$
\eqalign{
[\Theta + \Psi](\eta_0,k,\mu) & = \int_0^{\eta_0}d\eta (\hat\Theta_0 + \Psi
- i\mu \hat V_b)\left\{ \Vg e^{-[k/k_D(\eta)]^2} \right\} e^{ik\mu(\eta-\eta_0)}
\cr
& \qquad + \int_0^{\eta_0} d\eta R\Psi(e^{-[k/k_D(\eta)]^2} - 1) \Vg
	e^{ik\mu(\eta-\eta_0)}.
}
\Ceqn\CeqnBoltzdamp
$$
Although baryon drag effects such as acoustic displacement and enhancement
are already encorporated in the acoustic solution $\hat\Theta_0$ 
[see equation \Adis\AeqnSimple],
the residual 
$R\Psi$ term appears beneath the diffusion scale.  Let us ignore this 
for the moment.
The {\it effective} visibility for the acoustic oscillations is given by
$$
\hat \Vg = \Vg e^{-[k/k_D(\eta)]^2}.
\Ceqn\CeqnVg
$$
This function is plotted for a given model in \Figno\dampvis.
Unlike $\Vg$, $\hat \Vg$ is exponentially damped at late times by the
growing diffusion length and thus peaks at earlier times.  Note that a $10\%$ shift
in redshift represents a factor of three in optical depth near last
scattering.  Thus the region where we expect the approximation to break 
down is given little weight in the integral.  More specifically, the 
exponential damping insures that most contributions come from before
the epoch at which the diffusion length surpasses the wavelength.  
As we have seen in Appendix A, the optical depth through a wavelength is
high at this time and justifies the tight coupling expansion.

The damping of acoustic modes through last scattering can in general occur
due to two different mechanisms working in equation \Cdis\CeqnBoltzdamp: 
diffusion and cancellation.   However the effects are of greatly unequal 
magnitude.  Cancellation occurs since on small scales many wavelengths
of the perturbation span the Compton visibility function.  Photons that
last scattered at the crests of the perturbation will have their effect
cancelled by those that scattered at the troughs.  Mathematically
this occurs in equation 
\Cdis\CeqnBoltzdamp\ because 
the oscillating plane wave is integrated over the 
visibility function.  Cancellation leads to 
a power law damping of fluctuations as the scale decreases below the
width of the visibility function.  However, in the case of diffusion damped
acoustic contributions, it is not the width of the Compton visibility function
$\Vg$ that is relevant but rather the acoustic visibility function 
$\hat \Vg$.  As one goes to smaller and smaller scales (high $k$), the 
width of this function {\it decreases} as well.  Thus even at high $k$
the cancellation regime is never fully reached and one may approximate
the integral \Cdis\CeqnBoltzdamp\ by replacing $\hat \Vg$ by a delta function,
\ie\
$$
[\Theta + \Psi](\eta_0,k,\mu) \approx [\hat\Theta_0 + \Psi
- i\mu \hat V_b](\eta_*,k) e^{ik\mu(\eta_*-\eta_0)}\Dg(k),
\Ceqn\CeqnBoltzdelta
$$
where 
$$
\Dg(k) = \int_0^{\eta_0} d\eta \hat \Vg = \int_0^{\eta_0} 
d\eta \Vg e^{-[k/k_D(\eta)]^2}.
\Ceqn\CeqnDg
$$
The observable anisotropy follows by decomposing 
equation  \Cdis\CeqnBoltzdelta\ into Legendre moments 
$\Theta = \sum (-i)^{\ell}\Theta_\ell P_\ell(\mu)$
and 
summing over $k$-modes $C_\ell = {2 \over \pi} \int 
k^3|\Theta_\ell(\eta_0,k)/(2\ell+1)|^2 d\ln k$ (HSa).  
Decoupling thus increases the effective diffusion damping
length due to the corresponding increase 
in the mean free path of the photons.  The result is
a near exponential damping with scale that completely overwhelms the small
residual cancellation damping.

\topinsert
\centerline{
\hskip -0.5truecm
\epsfxsize=3.0in \epsfbox{rpsilsa.epsf} \hskip 0.5 truecm
\epsfxsize=3.0in \epsfbox{rpsilsb.epsf}
}

\baselineskip=12truept \leftskip=3truepc \rightskip=3truepc 
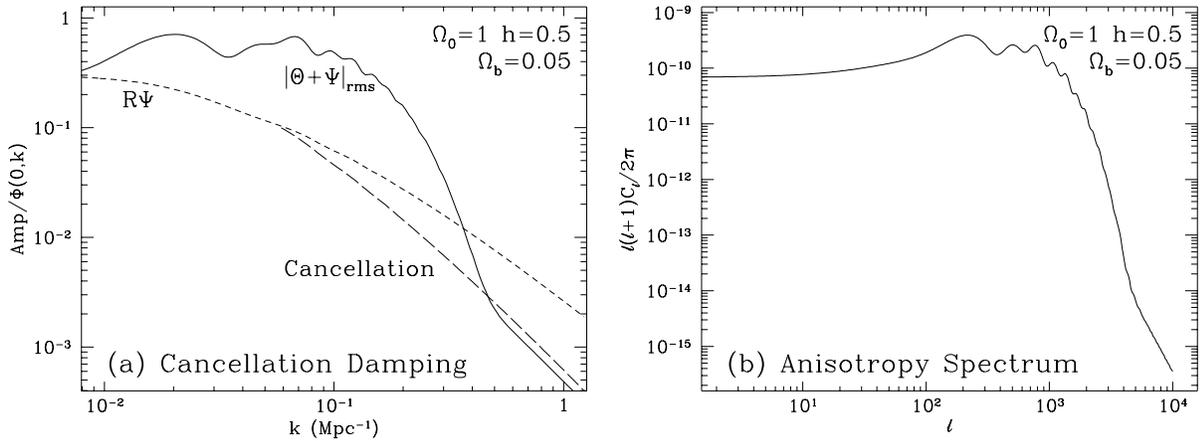
\noindent {\bf \Figno\rpsils.}  The residual baryon drag effect after
last scattering in an adiabatic CDM model. 
On scales under the width of the visibility 
function, cancellation between contributions
which came from potential wells and hills at last scattering
damps fluctuations from the baryon drag effect.  Note that
cancellation damping is weak and scales as 
$(k\eta_*)^{-1/2}$ in contrast to the exponential diffusion
damping. Projection relates the rms fluctuation in (a) to the
anisotropy power spectrum in (b).  
\endinsert
Cancellation damping does occur for the baryon drag effect $-R \Psi$ in
equation \Cdis\CeqnBoltzdamp\ which remains after diffusion damping.
The amplitude of the resultant fluctuations
can be estimated by noting that 
$$
[\Theta + \Psi](\eta_0,k,\mu)  =  -
	\int_0^{\eta_0} d\eta \Vg R\Psi e^{ik\mu(\eta-\eta_0)} 
\Ceqn\CeqnRPsi
$$
is approximately a Fourier transform.  Employing Parceval's theorem, we
obtain
$$
\eqalign{
|\Theta_0 + \Psi|^2_{\rm rms} & \equiv {1 \over 2}
	\int_{-1}^{1} |\Theta + \Psi|^2 d\mu \cr
&\approx {\pi \over k} \int_0^{\eta_0} |R\Psi \Vg|^2 d\eta \cr
&\approx {\pi \over k} |\Psi(\eta_*,k)|^2 \int_0^{\eta_0} (R 
	\Vg)^2 d\eta.
}
\Ceqn\CeqnParceval
$$
Thus the contribution to the rms is suppressed by roughly $(k\eta_*)^{-1/2}$
due to cancellation. 
Note that the residual baryon drag effect only appears in models where $\Psi$
itself is not damped away by diffusion, \eg\ adiabatic CDM and isocurvature
BDM models.
 In \Figno\rpsils a, we display this effect.  
The amplitude of the effect is slightly overestimated since baryon drag
weakens through last scattering.
Its effect on the anisotropy is shown in \Figno\rpsils b and 
can be obtained analytically in a computationally
simple matter through the approximations of Hu \& White (1995). Since it is unlikely
to be observable due to effects in the foreground of last scattering, we 
omit a detailed calculation here.

\goodbreak
\subsection{C.2 Baryon Decoupling}

The formal solution to the baryon Euler equation \Adis\AeqnBaryon\ is 
$$
a V_b = \int_0^{\eta}  d\eta' \, a [\dot \tau_d \Theta_1
	+ k\Psi] e^{-\tau_d} ,
\Ceqn\CeqnBaryonFormal
$$
where recall $\dot\tau_d = \dot \tau/R$ and quantities in the
integrand
are evaluated at $\eta'$.  
The two source terms are the Compton
drag effect
and infall into potential wells.  
The former forces the baryon velocity to follow the photon dipole (velocity)
at high drag optical depth $\tau_d$.
The
presence of the scale factor $a$ in the equation represents the fact that 
baryon velocities decay as $a^{-1}$ in the absence of sources.  
Since $\dot \tau_d e^{-\tau_d}$ is very nearly a delta function with respect
to variations on the expansion time scale, this equation is conceptually 
identical to its photon analogue 
\Cdis\CeqnBoltzformal\ with the replacement 
$\dot \tau \rightarrow \dot \tau_d$.  The plane wave factor is absent 
for the baryons since their particle velocities are low and the streaming
can be neglected in comparison to the wavelength.  
Drag depth unity $\tau_d(z_d) = 1$ marks the transition between
the drag and infall epochs.  For $\Omega_b h^2 \simgt 0.03$, the drag
epoch precedes last scattering $z_d > z_*$ assuming standard 
recombination.  
If the universe is reionized after recombination to some 
ionization level $x_e$, $z_d = 263 (\Omega_0 h^2)^{1/5} x_e^{-2/5} 
(1-Y_p/2)^{-2/5} \Theta_{2.7}^{-8/5}$ and the drag epoch significantly
precedes last scattering in most 
scenarios.\footnote{$^{\rm \dag}$}{This differs from the
treatment of (HSb) where $z_d$ was defined to be the epoch when 
the perturbation joined the growing mode of pressureless linear theory.
The presence of a decaying mode lowers this redshift by a factor of
$3 \over 5$ [see Appendix D, eqn.~(D-19)].}
Here $Y_p$ is the helium mass
fraction $Y_p \approx 0.23$

By analogy to the photon case, it is useful to define the
drag visibility
function
$$
\Vb = {a \dot \tau_d e^{-\tau_d} \over
       \int_0^{\eta_0}d\eta a \dot \tau_d e^{-\tau_d} },
\Ceqn\CeqnVb
$$
suitably normalized to have unity area.
Diffusion damping modifies the acoustic visibility function as
$$
\hat \Vb = \Vb e^{-[k/k_D(\eta)]^2}.
\Ceqn\CeqnhatVb
$$
Thus, we expect that immediately after the drag epoch the baryon 
velocity and density perturbations are approximately
$$
\eqalign{
V_b(\eta_d,k) & = \hat \Theta_1(\eta_d,k) \Db(k), \cr
\delta_b(\eta_d,k) & = \hat \delta_b(\eta_d,k) \Db(k), }
\Ceqn\CeqnAcousticVb
$$
where
$$
\Db(k) = \int_0^{\eta_0} d\eta \hat \Vb 
       = \int_0^{\eta_0} d\eta \Vb e^{-[k/k_D(\eta)]^2},
\Ceqn\CeqnDb
$$
and recall that $\hat \Theta_1$ and $\hat \delta_b$ were given in
\Bdis\BeqnAcoustic.

\topinsert
\centerline{
\epsfxsize=3.5in \epsfbox{cdmexact.epsf}
}

\baselineskip=12truept \leftskip=3truepc \rightskip=3truepc 
\noindent {\bf \Figno\cdmexact.}
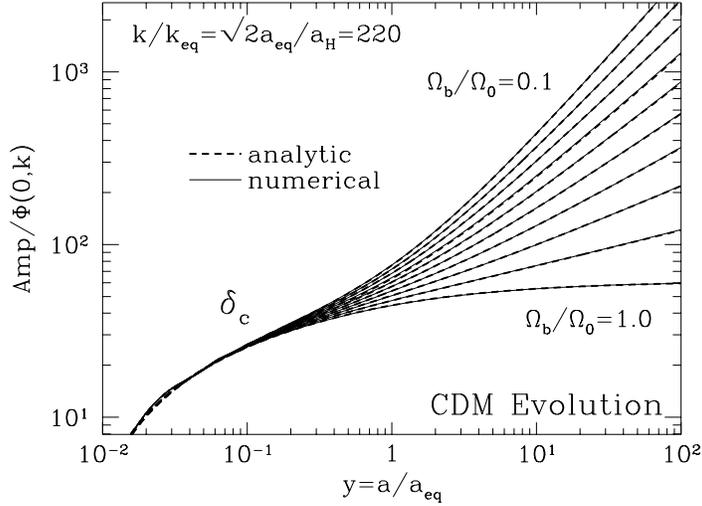
CDM evolution in the Compton drag epoch.  
If baryons contribute a significant fraction
of the total matter density, CDM growth will be slowed between equality
and the drag epoch.  Held by Compton drag, the baryons do not contribute
their self-gravity. For the numerical results, we choose a model that
never recombined so that $a_d \gg a_{eq}$.
\endinsert
\section{Appendix D. Matter Evolution}

After horizon crossing but 
before the end of the drag epoch, baryons 
follow the acoustic solution of
\Cdis\CeqnAcousticVb\ and the CDM follows their own
 pressureless evolution. 
After the drag epoch, the baryon evolution equation \Adis\AeqnBaryon\ is
identical to the cold dark matter and their joint evolution can be expressed
in terms of fluctuations in the total non-relativistic matter 
density $\delta_m$.   Thus, the solution for the time evolution of the
matter fluctuations requires knowledge of both the baryon and CDM perturbations
at the drag epoch. 
The baryonic contribution was obtained in Appendix C. Let us now evaluate
the CDM contributions.
\goodbreak
\subsection{D.1 Exact CDM Solutions}

The evolution of CDM fluctuations is described by equation \Bdis\BeqnCDM.
Since the curvature or $\Lambda$ terms are negligible before the drag epoch, 
this equation can be rewritten in terms of the equality-normalized scale factor
$y = a/a_{eq}$ as 
$$
{d^2 \over dy^2}\delta_c 
+{(2+3y) \over 2y(1+y)}{d \over dy}\delta_c =
{3 \over 2 y(1+y)}{\Omega_c \over \Omega_0}\delta_c \, .
\Eeqn\Eeqndeltacy
$$
Here we have assumed that 
the radiation contributions to the gravitational potential have
decayed to zero well after horizon crossing.  
In typical adiabatic models, the CDM contribution
usually dominates the non-relativistic matter.  Consider first the limit 
of negligible baryon fraction, $\Omega_b /\Omega_0 \ll 1$.
In this case, the matching condition at the drag epoch becomes trivial
since the baryons have no effect on the CDM. 
If $\Omega_c = \Omega_0$,  equation \Edis\Eeqndeltacy\ has same solution
before and after the drag epoch 
(see Peebles 1980 eqns. 12.5, 12.9),
$$
\eqalign{
U_1 = & {2 \over 3} + y \, , \cr
U_2 = & {15 \over 8}(2 + 3y)\ln\left[
{ (1+y)^{1/2} + 1 \over (1+y)^{1/2} - 1}\right] - 
{45 \over 4} (1+y)^{1/2} \, , \cr
}
\Eeqn\EeqnU
$$
before curvature or $\Lambda$ domination.  
Matching to the radiation dominated solution \Bdis\BeqnCDMrd, we obtain
$$
\delta_T(\eta,k)\simeq\delta_c(\eta,k)= I_1\Phi(0,k)
\left[{3 \over 2}\ln\left( 4 I_2 e^{-3}{a_{eq} \over a_H}\right)
U_1(\eta)
-{4 \over 15}U_2(\eta)\right] \, ,
\Eeqn\EeqnECasea
$$
for $k \gg k_{eq}$.

\smallskip

Let us now 
solve the equation \Edis\Eeqndeltacy\ for 
the case that the contribution of 
baryon is not negligible.
The two independent solutions are given in exact form through 
Gauss' hypergeometric function $F$ by 
$$
U_i =  (1 + y)^{-\alpha_i} F(\alpha_i,\alpha_i+{1 \over 2},
2\alpha_i + {1 \over 2} ; {1 \over 1 + y}) \, ,
\Eeqn\EeqnUGeneral
$$
where $i=1, 2$ and 
$$
\alpha_i = {1 \pm \sqrt{1 +24 \Omega_c/\Omega_0} \over 4} \, ,
\Eeqn\EeqnPower
$$
with $-$ and $+$ for $i=1$ and $2$, respectively.  Note that 
$\lim_{y \rightarrow \infty} U_i = y^{-\alpha_i}$.  Thus the main effect of
the baryons is to slow the power law growth of CDM after equality.

It is easy show that these solutions are identical to equations 
\Edis\EeqnU\
for $\Omega_c=\Omega_0$.  They also take on elementary forms for
two other special cases: 
$\Omega_c =0$, 
$$
\eqalign{
U_1 = & 1 \, , \cr
U_2 = & {1 \over 2}\ln\left[
{ (1+y)^{1/2} + 1 \over (1+y)^{1/2} - 1}\right] \, , \cr 
}
\Eeqn\EeqnUCaseb
$$
and $\Omega_c = {1 \over 3}\Omega_0$,
$$
\eqalign{
U_1 = & (1+y)^{1/2} \, , \cr
U_2 = & {3 \over 2} (1+y)^{1/2}\ln\left[
{ (1+y)^{1/2} + 1 \over (1+y)^{1/2} - 1}\right] -3 \, . \cr
}
\Eeqn\EeqnUCasec
$$

In order to map the solution for the radiation dominated limit 
\Bdis\BeqnCDMrd\ onto amplitudes of $U_1$ and $U_2$, 
we have to take the limit as $y \rightarrow 0$ of \Edis\EeqnUGeneral.
By using a linear transformation 
of the  hypergeometric function (see e.g. \ref\Abramowitz [\Abramowitz]
eq. 15.3.9),
we find
$$
\lim_{y \rightarrow 0}U_i = 
{\Gamma(2\alpha_i+1/2) \over \Gamma(\alpha_i)\Gamma(\alpha_i+1/2)}
\left[ -\ln{y} +2\psi(1) - \psi(\alpha_i) -\psi(\alpha_i+1/2) \right]
 \, ,
\Eeqn\EeqnULimit
$$
where $\Gamma(x)$ and 
$\psi(x)
 =\Gamma'(x)/\Gamma(x)$
are gamma and digamma functions respectively.
Matching to the radiation dominated solution \Bdis\BeqnCDMrd\ yields
$$
\delta_c (\eta,k) = I_1\Phi(0,k)
\left[A_1 U_1(\eta) + A_2 U_2(\eta) \right]\, ,
\Eeqn\EeqnEvolForm
$$ 
where 
$$
\eqalign{
A_1 = & - {\Gamma(\alpha_1)\Gamma(\alpha_1+1/2)\over \Gamma(2\alpha_1+1/2)
\left[ \psi(\alpha_1)+\psi(\alpha_1+1/2)
-\psi(\alpha_2)-\psi(\alpha_2+1/2)\right]} \cr
& \times\left[ \ln{\left(I_2{a_{eq}\over a_H}\right)} 
+2\psi(1) - \psi(\alpha_2) -\psi(\alpha_2+1/2) \right]
\, . \cr
}
\Eeqn\EeqnAExact
$$ 
$A_2$ is obtained by replacing the subscripts $1 \leftrightarrow 2$ of $A_1$.
Since $\Omega_b / \Omega_0 \le 1$, 
it is useful to approximate the coefficients
with a series expansion,
$$
A_1 = B_1 \ln \left( I_2 {a_{eq} \over a_H} \right) + B_2,
\Eeqn\EeqnAApprox
$$ 
with
$$
\eqalign{
B_1 & = {3 \over 2} \left[ 1 - 0.568 (\Omega_b/\Omega_0) 
   + 0.094 (\Omega_b/\Omega_0)^2 + {\cal O}[(\Omega_b/\Omega_0)^3]\right], \cr
B_2 & = {3 \over 2}(\ln 4 - 3) \left[ 1 - 1.156 (\Omega_b/\Omega_0) 
   + 0.149 (\Omega_b/\Omega_0)^2 - 0.074(\Omega_b/\Omega_0)^3  
   +  {\cal O}[(\Omega_b/\Omega_0)^4] \right],
}
\Eeqn\EeqnB
$$
valid at the percent level for $\Omega_b/\Omega_0 < 1/2$, and
$$
A_2 = - {\Gamma(\alpha_2)\Gamma[\alpha_2+1/2] \over \Gamma[2\alpha_2 + 1/2]}
	\left[ {\Gamma(2\alpha_1 + 1/2)  \over \Gamma(\alpha_1) \Gamma(\alpha_1
	+ 1/2) } A_1 + 1 \right].
\Eeqn\EeqnAAppproxb
$$
For the three special cases, we can describe $\delta_c$ by elementary
functions.  If $\Omega_c = \Omega_0$, $\delta_c$ is given by 
equation \Edis\EeqnU, whereas
$$
\eqalign{
\delta_c (\eta,k) = & I_1\Phi(0,k)
\left[ 
\ln{\left( 4I_2{a_{eq} \over a_H} \right)}
U_1 -  2 U_2 \right]  \qquad \Omega_c=0  \, , \cr
\delta_c (\eta,k) = & I_1\Phi(0,k)
\left[ \ln{\left( 4I_2 e^{-2}{a_{eq} \over a_H} \right)}
U_1 -  {2 \over 3} U_2 \right]
   \qquad \Omega_c={1 \over 3}\Omega_0  \, . \cr
}
\Eeqn\EeqnUCases
$$
In \Figno\cdmexact, we 
show the time evolution of $\delta_c$ before the drag epoch
for several different values of $\Omega_b/\Omega_c$.  Numerical
results in this figure are for fully reionized models so that the
drag epoch ends well after equality unlike other examples in this
paper.

\topinsert
\centerline{
\epsfxsize=3.5in \epsfbox{velov.epsf}
}

\baselineskip=12truept \leftskip=3truepc \rightskip=3truepc 
\noindent {\bf \Figno\velov.} 
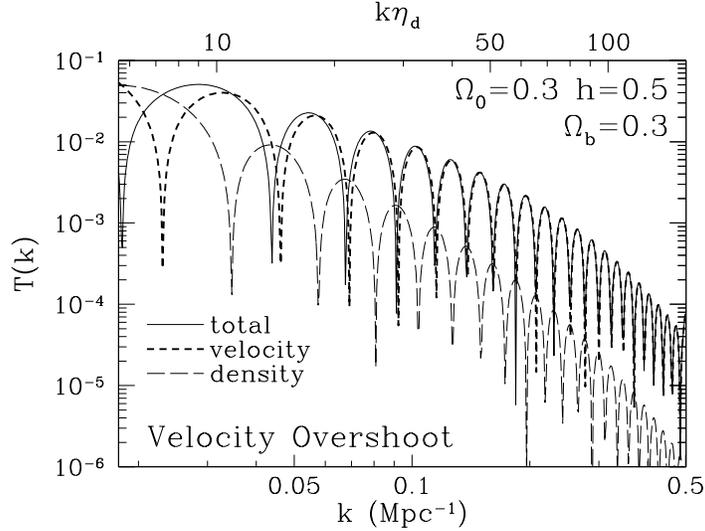
Velocity overshoot effect.  Below the horizon at the drag epoch
$k\eta_d \gg 1$, the acoustic velocity at $z_d$ dominates the growing
mode and hence the final transfer function.  Near the horizon, the
acoustic density becomes comparable and shifts the zero points of
the oscillation. 
\endinsert
\subsection{D.2 Matter Transfer Function}

With the baryon and CDM fluctuations at the drag epoch from equations 
\Cdis\CeqnAcousticVb\ and
\Edis\EeqnEvolForm\ respectively, 
we can now solve for the evolution to the present.  
After the drag
epoch, baryons behave dynamically as CDM and the combined 
non-relativistic matter fluctuations 
$$
\eqalign{
\delta_m & = {\Omega_b \over \Omega_0} \delta_b
			+ \left(1 - {\Omega_b \over \Omega_0} \right) 
			\delta_c,
			\cr
V_m & = {\Omega_b \over \Omega_0} V_b
			+ \left(1 - {\Omega_b \over \Omega_0} \right) 
			V_c.
}
\Eeqn\EeqnComponent
$$
follow the growing and decaying solutions for $\delta_m$ (Peebles 1980)
$$
\eqalign{
D_1 = & {2 \over 3} + y \, , \cr
D_2 = & {15 \over 8}(2 + 3y)\ln\left[
{ (1+y)^{1/2} + 1 \over (1+y)^{1/2} - 1}\right] - 
{45 \over 4} (1+y)^{1/2} \, , \cr
}
\Eeqn\EeqnD
$$
before curvature or $\Lambda$ domination.  To account for 
effects from curvature and $\Lambda$ at $a \gg a_{eq}$ or $y \gg 1$,
one simply needs to replace $a \rightarrow D$, where
$$
\eqalign{
D(a) & = {5 \over 2} \Omega_0 g(a) \int {d a' \over [a'g(a')]^{3}}, \cr
g^2(a) &= a^{-3}\Omega_0 + a^{-2} (1-\Omega_0 -\Omega_\Lambda) + \Omega_\Lambda,
\cr
}
\Eeqn\EeqnDNormal
$$
is the growing mode of radiationless linear theory normalized to equal 
$a$ at early times.

By matching the fluctuations at the drag epoch, we obtain for the growing 
mode
$$
\eqalign{
\delta_m(\eta,k) & = [ G_1(\eta_d) \delta_m(\eta_d,k) + 
		       G_2(\eta_d) k V_m(\eta_d,k)]  D_1(\eta),\cr
G_1(\eta) & = {\dot D_2 \over D_1 \dot D_2 - \dot D_1 D_2}, \qquad 
G_2(\eta) = {D_2 \over D_1 \dot D_2 - \dot D_1 D_2} \cr
}
\Eeqn\EeqnMatching
$$
and similarly for the decaying mode.
The combined time evolution is plotted in \Figno\bartime\ for a BDM model
with $\delta_m = \delta_b$.
If $z_d \ll z_{eq}$, equation \Edis\EeqnMatching\ 
reduces to the familiar form
$$
\delta_m(\eta,k) = {a \over a_d} \left[ {3 \over 5} \delta_m(\eta_d,k)
 - {1 \over 5} (k\eta_d) V_m(\eta_d,k) \right].
\Eeqn\EeqnMDMatching
$$
Notice that on scales much less than the horizon at the drag epoch, the
 velocity at $\eta_d$ dominates the growing mode if the two values
are comparable at $\eta_d$ (see \Figno\velov).  
This ``velocity overshoot'' effect occurs since the
peculiar velocity moves the matter and creates density fluctuations
kinematically. Expansion drag on the velocity eliminates it in
an expansion time $\eta_d$ and thus causality prevents this effect
from generating density fluctuation above the horizon at the drag epoch 
$k\eta_d \ll 1$. 

It is conventional to recast the evolutionary effects in
terms of a transfer function.  As with equation \Edis\EeqnComponent,
we can break the present day transfer function 
up into a baryonic and cold dark matter contribution
at the drag epoch
$$
T(k) = {\Omega_b \over \Omega_0} T_b(k) +
\left(1 - {\Omega_b \over \Omega_0}\right) T_c(k).
\Eeqn\EeqnTparts
$$
It should be kept in mind that $T_b$ and $T_c$ do {\it not}
represent the respective transfer functions today.
Let us consider the adiabatic transfer function.   Here one expresses
the evolution of 
small scale fluctuations in terms of
those at large scales, \ie\
$|\delta_T(\eta_0,k)|^2 \propto T^2(k) P(k)$ with normalization
$\lim_{k \rightarrow 0} T(k) = 1$ and initial 
power spectrum $P(k) \propto |k^2 \Phi(0,k) |^2 $. 
The large scale solution is given by
(see HSa eqn.~A-15)
$$
\lim_{k \rightarrow 0} \delta_T(\eta,k) = \left( 1 + {4 \over 15} f_\nu
	\right) \left( 1 + {2 \over 5} f_\nu \right)^{-1} {6 \over 5} 
	\left( {k \over k_{eq}} \right)^2 D_1(\eta) \Phi(0,k).
\Eeqn\EeqndeltacLS
$$
The acoustic contribution of the baryons is therefore,
$$
T_b(k) = {15 \over 4}
	 \left(  1 + {4 \over 15} f_\nu   \right)^{-1}
        \left( {k_{eq} \over k} \right)^2  \Db(k) 
 (1+R)^{-1/4}
	\left( \cos kr_s
	  - {D_2 \over \dot D_2 }  
	  k c_s \sin kr_s
	\right) G_1 \Bigg|_{\eta=\eta_d} . 
\Eeqn\EeqnTba
$$
This equation is compared with numerical results for the
adiabatic transfer function in \Figno\trans a.

For isocurvature BDM models, $T = T_b$ and
it is conventional to define it such that $|\delta_T(\eta_0,k)|^2 \propto
T^2(k) | S(0,k) |^2$ with normalization $\lim_{k \rightarrow \infty} 
T(k) = 1$.   From equation \Bdis\BeqnAmpIsocurvature, 
the small scale tail of the transfer function becomes
$$
T(k)  = 1 - {3\sqrt{6} \over 4}
	 \left(  1 - {4 \over 15} f_\nu   \right)
        {k_{eq} \over k} \Db(k)  (1+R)^{-1/4}
	\left( \sin kr_s
	  - {D_2 \over \dot D_2 }  
	 k c_s \cos k r_s\right) \Bigg|_{\eta=\eta_d}.
\Eeqn\EeqnTbi
$$
This function is plotted in \Figno\trans b.

In the $\Omega_b/\Omega_0 \rightarrow 0$ limit, the 
CDM contributions can be expressed in terms of elementary functions
as 
$$
\eqalign{
\lim_{\Omega_b \rightarrow 0} T(k) &= I_1 \left( 1 + {2 \over 5} f_\nu \right) 
 	\left(1+{4 \over 15}f_\nu\right)^{-1} {5 \over 4} \left({k_{eq} \over k}
	\right)^2 \ln \left( 4I_2 e^{-3} {a_{eq} \over a_H} \right) 
	\cr
     &\approx {\ln 1.8 q \over 14.2 q^2}, \qquad f_\nu = 0.405
}
\Eeqn\EeqnAnalyticT
$$
where $q = (k/{\rm Mpc}^{-1}) \Theta_{2.7}^2 (\Omega_0 h^2)^{-1}$.  
This should be
compared with the high $k$ tail of the standard fitting function 
to the numerical results (BBKS),
$$
T_{\rm BBKS}(q) = 
{\ln(1 + 2.34q) \over 2.34 q} [ 1 + 3.89 q + (16.1 q)^2 + 
(5.46q)^3 + (6.71 q)^4]^{-1/4},
\Eeqn\EeqnBBKS
$$
\ie\ $\lim_{q \rightarrow \infty} T(q) = \ln (2.34 q)/ 15.7 q^2$ which 
differs by $\sim 10\%$ from the analytic prediction at small scales.  In
fact, since the fitting formula was designed to fit intermediate
scales, equation \Edis\EeqnAnalyticT\ is more accurate at extremely 
small scales.

If the baryon fraction is non-negligible, the contribution is expressed
in terms of hypergeometric functions through equation \Edis\EeqnUGeneral,
$$
T_c  = I_1\left( 1 + {2 \over 5} f_\nu \right) 
 	\left(1+{4 \over 15}f_\nu\right)^{-1} {5 \over 6} \left({k_{eq} \over k}
	\right)^{2} 
    \left\{ G_1 [A_1 U_1 + A_2 U_2]
	- G_2 [A_1 \dot U_1 + A_2 \dot U_2] \right\} \Big|_{\eta=\eta_d}.
\Eeqn\EeqnCDMTransMessy
$$
Though exact, this expression is rather complicated.  
It is useful and instructive to seek a simple scaling relation for 
this form.  Notice that for all cases the scale dependence of
the transfer function may be written as
$$
T_c \approx \alpha {\ln 1.8 \beta q  \over 14.2 q^2} ,
\Eeqn\EeqnScaling
$$
for $f_\nu = 0.405$.  Here $\alpha$ and $\beta$ are functions
of $\Omega_0 h^2$ and $\Omega_b/\Omega_0$.  Note that 
as $q \rightarrow \infty$ the
modification due to $\beta$ becomes insignificant.  
A very accurate fit to both $\alpha$ and $\beta$ is given in Appendix E
eqns. (E-12) (E-13).
The numerical, analytic and fitted analytic results are compared with 
the empirical scalings of (Peacock \& Dodds 1994, Sugiyama 1995) 
in \Figno\transcdm.
The analytic calculation is essentially exact while the fitted analytic
form works to 1\% accuracy.
Notice that in this extreme case we have significantly improved upon
previous results.

Equation \Edis\EeqnScaling\ breaks down for intermediate
to large scales. The CDM contribution can be approximately
scaled from the
BBKS form as
$$
\eqalign{
T_c(k) & \approx T_{\rm BBKS}(\tilde q),\cr
\tilde q(k) &  = {k \over {\rm Mpc^{-1}}} \alpha^{-1/2}
	(\Omega_0 h^2)^{-1} \Theta_{2.7}^2.
}
\Eeqn\EeqnTc
$$
This expression is employed in \Figno\trans\ with
the coefficient $6.71$ in \Edis\EeqnBBKS\ replaced by $6.71 (14.2/15.7)
= 6.07$ to match the analytic small scale tail.
Notice that at the largest scales, this expression underestimates the matter
transfer function.  This is because 
baryon contributions must be properly included.
Although the limiting form $\lim_{k\rightarrow 0} T_b  = 1$ is simple,
the behavior near the horizon scale at $z_d$ is not.   Since this region 
is not the main focus of this work, we do not attempt to describe this
analytically.  If the baryonic oscillations are small or smoothed over,
an approximate patch is given by
$$
\eqalign{
T(k) & \approx T_{\rm BBKS}(\hat q),\cr
\hat q(k) & = {k \over {\rm Mpc^{-1}}} 
	\left( 1 -{\Omega_b \over \Omega_0} \right)^{-1/2}
	\alpha^{-1/2} (\Omega_0 h^2)^{-1} \Theta_{2.7}^2,
}
\Eeqn\EeqnTcapprox
$$
which extends the Peacock \& Dodds (1994) approach to high 
$\Omega_b/\Omega_0$.

\topinsert
\centerline{
\epsfxsize=3.5in \epsfbox{trans.epsf}
}

\baselineskip=12truept \leftskip=3truepc \rightskip=3truepc 
\noindent {\bf \Figno\transcdm.}
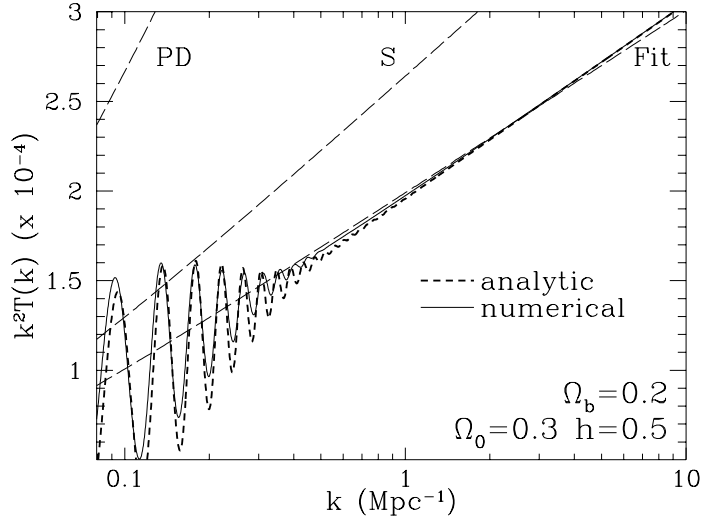
Adiabatic CDM transfer function in a high $\Omega_b/\Omega_0=2/3$ case.
The analytic solution is essentially exact 
in the small scale limit.  Simple fits based on the BBKS form
can cause large errors at the small scale: PD (Peacock \& Dodds 1994) 
and S (Sugiyama 1995).
The fitting function developed here [see equation
\Edis\EeqnScaling] works at the $1\%$ level even for this extreme case. 
\endinsert
Finally, we can express the ratio of the acoustic peak heights to the 
CDM tail with equations \Edis\EeqnTparts, \Edis\EeqnTba\ and 
\Edis\EeqnAnalyticT,
$$
{\Omega_b \over \Omega_c} {T_b \over T_c} \approx  {\sqrt{3}\over I_1}  
 k G_2(\eta_d) [1+R(\eta_d)]^{-3/4} \Db(k) \left(1 + {2 \over 5} f_\nu 
	\right)^{-1} \left[ 
	\ln \left( {4I_2 e^{-3}} {a_{eq} \over a_H}\right) \right]^{-1} 
	{\lim_{\Omega_b \rightarrow 0} T_c \over T_c}
\Eeqn\EeqnTRatio
$$
if the velocity overshoot effect dominates the acoustic contributions.
We can simplify this expression by noting that $G_2(\eta_d) \approx
{2 \over 5} (\Omega_0 H_0^2 /a_{eq})^{-1/2} (1+a_d/a_{eq})^{-1/2}$.
Furthermore, the function $k \Db$ peaks at roughly $0.8 k_S$ with 
an amplitude of $0.4 k_S$.  
With the scaling of equation \Edis\EeqnScaling, 
the peak relative amplitude of the acoustic oscillation is approximately
$$
70 {\Omega_b \over \Omega_c}
	{k_S \over {\rm Mpc}^{-1}} 
	{a_{eq} \over (a_{eq}+a_d)^{1/2}}
	[1 + R(\eta_d)]^{-3/4}
	(\Omega_0 h^2)^{-1/2} 
	\alpha^{-1}
	\left[\ln\left({1.8\beta\over \Omega_0 h^2}
	{k_S \over {\rm Mpc}^{-1}} \right)\right]^{-1}.
\Eeqn\EeqnTransRatio
$$
The logarithmic term is roughly unity and may be dropped for estimation 
purposes [{\it cf.} equation \eqnTransRatio]. 
Equation \Edis\EeqnTransRatio\ roughly quantifies the prominence of 
the acoustic oscillations in a CDM model.  For best accuracy however, 
the solutions 
\Edis\EeqnTba\  and
\Edis\EeqnCDMTransMessy\ 
for the baryons and CDM respectively should be employed.  

\section{Appendix E: Recombination Fitting Formulae}

Rather than recalculate the atomic physics of
recombination each time one needs
to consider effects at the last scattering and drag epochs, it is 
convenient to have accurate fitting formulae that encorporate the 
ionization history.  In general, all quantities associated with the
ionization history must be functions of $\Omega_0 h^2$ and $\Omega_b h^2$ 
alone once the CMB temperature $T_0 = 2.726$K (Mather \etal\ 1994),
neutrino fraction $f_\nu = 0.405$, and helium fraction $Y_p \approx 0.23$
are fixed.  Fitting functions
in this Appendix are designed to be valid at the percent level 
for an extended range of parameter space, $0.0025 \simlt 
\Omega_b h^2 \simlt 0.25$ and $0.025 \simlt \Omega_0 h^2 \simlt
0.64$ and consequently appear rather complicated.  We employ
a recombination calculation based on the improvements discussed in
Hu \etal\ (1995).

The last scattering epoch is a very weak function of parameters 
and is given by 
$$
\eqalign{
z_* & = 1048 [1 + 0.00124 (\Omega_b h^2)^{-0.738}]
        [1 + g_1 (\Omega_0 h^2)^{g_2} ] ,\cr
g_1 & = 0.0783 (\Omega_b h^2)^{-0.238} [1+39.5(\Omega_b h^2)^{0.763}]^{-1} ,\cr
g_2 & = 0.560 [1+21.1(\Omega_b h^2)^{1.81}]^{-1} .\cr
}
\Deqn\DeqnLS
$$
The drag epoch ends at a related redshift which depends 
somewhat more strongly on the parameters
$$
\eqalign{
z_d & = 1345 {(\Omega_0 h^2)^{0.251}
\over 1 + 0.659 (\Omega_0 h^2)^{0.828} }
[1 + b_1 (\Omega_b h^2)^{b_2}], \cr
b_1 & = 0.313 (\Omega_0 h^2)^{-0.419} [1 + 0.607 (\Omega_0 h^2)^{0.674} ], \cr
b_2 & = 0.238 (\Omega_0 h^2)^{0.223}. \cr
}
\Deqn\DeqnZdrag
$$
The two are approximately equal if $\Omega_b h^2 \approx 0.03$.  

The diffusion damping envelope can be approximated through the first
decade of damping by the form 
$$
\Dg(k) \approx e^{-[k/k_{D\gamma}]^{m_\gamma}},
\Deqn\DeqnGammaDampApp
$$
where the effective damping scale $k_{D \gamma}$ has simple asymptotic scaling,
$$
{k_{D\gamma} \over {\rm Mpc}^{-1}}
= \cases{ F_1, & ${\Omega_b h^2 \gg 0.1}$ \cr
                   (\Omega_b h^2)^{p_1}
                   F_2, & ${\Omega_b h^2
                     \ll 0.1}$\cr }
\Deqn\DeqnKgammaAsy
$$
with
$$
\eqalign{
F_1 & = 0.293 (\Omega_0 h^2)^{0.545}[1 + (25.1\Omega_0 h^2)^{-0.648} ], \cr
F_2 & = 0.524 (\Omega_0 h^2)^{0.505}[1 + (10.5 \Omega_0 h^2)^{-0.564} ], \cr
p_1 & = 0.29.
}
\Deqn\DeqnKgammafuncs
$$
The basic scaling in the low $\Omega_b h^2$ limit can be understood by
the Saha approximation in which the ionization fraction approximately
scales as $x_e  \propto (\Omega_b h^2)^{-1/2}$.  Thus the diffusion
length $\lambda_D \sim k_\gamma^{-1} \sim \sqrt{\eta_*/\dot\tau}
\propto  \eta_*^{1/2} (\Omega_b h^2)^{-1/4}$.  Since $\eta_*
\propto (\Omega_0 h^2)^{-1/2} $ in the matter dominated
high $\Omega_0 h^2$ limit, this is approximately of the same form
as equation \Ddis\DeqnKgammafuncs.
For high $\Omega_b h^2$, the corrections
from an accurate treatment of the atomic levels becomes more important
due to the high Lyman-$\alpha$ opacity.
These two simple limits can be accurately joined by a rather artificial
looking but highly accurate form
$$
\eqalign{
{k_{D\gamma} \over {\rm Mpc}^{-1}}
          & = \left\{ {2 \over \pi} {\rm arctan}\left[ {\pi \over 2}
           (F_2/F_1)^{p_2/p_1} (\Omega_b h^2)^{p_2}
        \right]\right\}^{p_1/p_2} F_1, \cr
p_2 & = 2.38 (\Omega_0 h^2)^{0.184} .
}
\Deqn\DeqnKgamma
$$
From \Figno\dampvis, we see that this overestimates the true damping 
as $\Omega_b h^2 \rightarrow 0$ due to a breakdown of tight coupling.  
It is interesting to note that the full numerical
results suggest that the Saha estimation of $p_1 \approx 0.25$ in
equations \Ddis\DeqnKgammafuncs\ and \Ddis\DeqnKgamma\ is a somewhat 
better phenomenological fit.

The steepness index $m_\gamma$ of the diffusion damping envelope is
a very weak function of cosmological parameters.  In the limit that
last scattering occurred instantaneously $m_\gamma \rightarrow 2$.  
The finite width of the visibility function modifies this as
$$
m_\gamma = 1.46 (\Omega_0 h^2)^{0.0303} \left(
	  1 + 0.128 {\rm arctan} 
	  \left\{ \ln[(32.8\Omega_b h^2)^{-0.643}] \right\} \right),
\Deqn\DeqnSlopegamma
$$
which only varies by $\sim 10\%$ across the full range of parameter space.

Silk damping for the baryons can likewise be approximated by 
$$
\Db(k)  \approx e^{-(k/k_S)^{m_b}}, 
\Deqn\DeqnDb
$$
with
$$
{k_S \over {\rm Mpc}^{-1}}
          = 1.38 (\Omega_0 h^2)^{0.398} (\Omega_b h^2)^{0.487}
        { 1 + (96.2\Omega_0 h^2)^{-0.684}
          \over
          1 + (346\Omega_b h^2)^{-0.842}},
\Deqn\DeqnSilk
$$
and the steepness index by
$$
m_S  = 1.40 {(\Omega_b h^2)^{-0.0297} (\Omega_0 h^2)^{0.0282}
        \over 1 + (781\Omega_b h^2)^{-0.926} }.
\Deqn\DeqnmS
$$
As is the case with the photons, the latter accounts for the
width of the visibility function and is almost independent of
cosmological parameters.  

Finally, by employing equation \Ddis\DeqnZdrag\ for the drag epoch,
the cumbersome analytic result for the 
CDM {\it drag}-contribution to the small scale transfer function 
$T = (\Omega_b/\Omega_0)T_b + (\Omega_c/\Omega_0)T_c$
from 
\Edis\EeqnCDMTransMessy\
can be fit as
$$
T_c \approx \alpha {\ln (1.8 \beta q) \over 14.2 q^2},
\Deqn\DeqnScaling
$$
with 
$$
\eqalign{
\alpha & = a_1^{-\Omega_b/\Omega_0} a_2^{-(\Omega_b/\Omega_0)^3}, \cr
a_1 & = (46.9 \Omega_0 h^2 )^{0.670} [1 + (32.1 \Omega_0 h^2)^{-0.532}], \cr 
a_2 & = (12.0 \Omega_0 h^2 )^{0.424} [1 + (45.0 \Omega_0 h^2)^{-0.582}],
}
\Deqn\Deqnalpha
$$
as the suppression factor and
$$ 
\eqalign{
\beta^{-1} & = 1 + b_1 [(\Omega_c/\Omega_0)^{b_2}-1],  \cr
b_1 & = 0.944 [ 1 + (458 \Omega_0 h^2)^{-0.708} ]^{-1}, \cr
b_2 & = (0.395 \Omega_0 h^2)^{-0.0266}, \cr	
}
\Deqn\Deqnbeta
$$
as the correction to the logarithm.

\eject

\vbox{ \vskip 20pt \centerline{
\vbox{ \offinterlineskip
\halign { \vrule#
& \quad # \hfil& \vrule#
& \quad # \hfil& \vrule#
& \quad # \hfil& \vrule# \cr
\noalign{\hrule} height2pt
&\omit& &\omit& &\omit&
\cr
& Symbol~& & Definition \quad&& Equation~& \cr
height2pt
&\omit& &\omit& &\omit&
\cr \noalign{\hrule}
height2pt
&\omit& &\omit& &\omit&
\cr
&$\Theta$                      &&CMB~temperature~pert.                                                 &&~\Adis\AeqnHierarchy                     & \cr
height2pt
&\omit& &\omit& &\omit&
\cr \noalign{\hrule}
height2pt
&\omit& &\omit& &\omit&
\cr
&$\Theta_0$                    &&CMB~monopole~pert.                                                    &&~\Adis\AeqnHierarchy                     & \cr
height2pt
&\omit& &\omit& &\omit&
\cr \noalign{\hrule}
height2pt
&\omit& &\omit& &\omit&
\cr
&$\Theta_1$                    &&$V_\gamma$--CMB~dipole~pert.                                          &&~\Adis\AeqnHierarchy                     & \cr
height2pt
&\omit& &\omit& &\omit&
\cr \noalign{\hrule}
height2pt
&\omit& &\omit& &\omit&
\cr
&$\Pi_T$                       &&Anisotropic~stress~pert.                                              &&~\Adis\AeqnPoisson                       & \cr
height2pt
&\omit& &\omit& &\omit&
\cr \noalign{\hrule}
height2pt
&\omit& &\omit& &\omit&
\cr
&$\Phi$                        &&Curvature~perturbation                                                &&~\Adis\AeqnPoisson                       & \cr
height2pt
&\omit& &\omit& &\omit&
\cr \noalign{\hrule}
height2pt
&\omit& &\omit& &\omit&
\cr
&$\Psi$                        &&Newtonian~potential                                                   &&~\Adis\AeqnPoisson                       & \cr
height2pt
&\omit& &\omit& &\omit&
\cr \noalign{\hrule}
height2pt
&\omit& &\omit& &\omit&
\cr
&$\Omega_i$                    &&Critical~fraction~in~$i$                                              &&~\eqnSound                               & \cr
height2pt
&\omit& &\omit& &\omit&
\cr \noalign{\hrule}
height2pt
&\omit& &\omit& &\omit&
\cr
&$\alpha$                      &&Baryon~$T$-suppression                                                &&~\Ddis\DeqnScaling                       & \cr
height2pt
&\omit& &\omit& &\omit&
\cr \noalign{\hrule}
height2pt
&\omit& &\omit& &\omit&
\cr
&$\alpha_1$~$\alpha_2$         &&$U_1$$U_2$~power~law                                                  &&~\Edis\EeqnPower                         & \cr
height2pt
&\omit& &\omit& &\omit&
\cr \noalign{\hrule}
height2pt
&\omit& &\omit& &\omit&
\cr
&$\beta$                       &&Baryon~$T$-log~correct.                                               &&~\Ddis\DeqnScaling                       & \cr
height2pt
&\omit& &\omit& &\omit&
\cr \noalign{\hrule}
height2pt
&\omit& &\omit& &\omit&
\cr
&$\eta~(\eta_t)$               &&Conformal~time~(at~t)                                                 &&~\Adis\AeqnHierarchy                     & \cr
height2pt
&\omit& &\omit& &\omit&
\cr \noalign{\hrule}
height2pt
&\omit& &\omit& &\omit&
\cr
&$\delta_i$                    &&Density~pert.~in~$i$                                                  &&~\eqnSuppression                         & \cr
height2pt
&\omit& &\omit& &\omit&
\cr \noalign{\hrule}
height2pt
&\omit& &\omit& &\omit&
\cr
&$\tau$                        &&Compton~optical~depth                                                 &&~\Adis\AeqnHierarchy                     & \cr
height2pt
&\omit& &\omit& &\omit&
\cr \noalign{\hrule}
height2pt
&\omit& &\omit& &\omit&
\cr
&$\tau_d$                      &&Drag~optical~depth                                                    &&~\Cdis\CeqnBaryonFormal                  & \cr
height2pt
&\omit& &\omit& &\omit&
\cr \noalign{\hrule}
height2pt
&\omit& &\omit& &\omit&
\cr
&$\omega$                      &&Acoustic~frequency                                                    &&~\Adis\AeqnDispersion                    & \cr
height2pt
&\omit& &\omit& &\omit&
\cr \noalign{\hrule}
height2pt
&\omit& &\omit& &\omit&
\cr
&${\cal~D}_\gamma$             &&Photon~damping~factor                                                  &&~\Cdis\CeqnDg                            & \cr
height2pt
&\omit& &\omit& &\omit&
\cr \noalign{\hrule}
height2pt
&\omit& &\omit& &\omit&
\cr
&${\cal~D}_b$                  &&Baryon~damping~factor                                                 &&~\Cdis\CeqnDb                            & \cr
height2pt
&\omit& &\omit& &\omit&
\cr \noalign{\hrule}
height2pt
&\omit& &\omit& &\omit&
\cr
&$\Vg$                         &&Compton~visibility                                                    &&~\Cdis\CeqnBoltzformal                   & \cr
height2pt
&\omit& &\omit& &\omit&
\cr \noalign{\hrule}
height2pt
&\omit& &\omit& &\omit&
\cr
&$\hat\Vg$                     &&CMB~acoustic~visibility                                               &&~\Cdis\CeqnVg                            & \cr
height2pt
&\omit& &\omit& &\omit&
\cr \noalign{\hrule}
height2pt
&\omit& &\omit& &\omit&
\cr
&$\Vb$                         &&Drag~visibility                                                       &&~\Cdis\CeqnVb                            & \cr
height2pt
&\omit& &\omit& &\omit&
\cr \noalign{\hrule}
height2pt
&\omit& &\omit& &\omit&
\cr
&$\hat\Vb$                     &&Baryon~acoustic~visibility                                            &&~\Cdis\CeqnAcousticVb                    & \cr
height2pt
&\omit& &\omit& &\omit&
\cr \noalign{\hrule}
height2pt
&\omit& &\omit& &\omit&
\cr
&$A_1~A_2$                     &&Boost~CDM~matching~cond.                                              &&~\Edis\EeqnAExact                        & \cr
height2pt
&\omit& &\omit& &\omit&
\cr \noalign{\hrule}
height2pt
&\omit& &\omit& &\omit&
\cr
&$C_A$                         &&Adiabatic~acoustic~amp.                                               &&~\Bdis\BeqnAmpAdiabatic                  & \cr
height2pt
&\omit& &\omit& &\omit&
\cr \noalign{\hrule}
height2pt
&\omit& &\omit& &\omit&
\cr
&$C_I$                         &&Isocurvature~acoustic~amp.                                            &&~\Bdis\BeqnAmpIsocurvature               & \cr
height2pt
&\omit& &\omit& &\omit&
\cr \noalign{\hrule}
height2pt
&\omit& &\omit& &\omit&
\cr
&$D$                           &&Radiationless~growth                                                  &&~\Edis\EeqnDNormal                       & \cr
height2pt
&\omit& &\omit& &\omit&
\cr \noalign{\hrule}
height2pt
&\omit& &\omit& &\omit&
\cr
&$D_1~D_2$                     &&Matter-radiation~growth                                               &&~\Edis\EeqnD                             & \cr
height2pt
&\omit& &\omit& &\omit&
\cr \noalign{\hrule}
height2pt
&\omit& &\omit& &\omit&
\cr
&$G_1~G_2$                     &&Drag~matching~cond.                                                   &&~\Edis\EeqnMatching                      & \cr
height2pt
&\omit& &\omit& &\omit&
\cr \noalign{\hrule}
height2pt
&\omit& &\omit& &\omit&
\cr
&$I_1~I_2$                     &&CDM~boost~integrals                                                   &&~\Bdis\BeqnCDMIntegrals                  & \cr
height2pt
&\omit& &\omit& &\omit&
\cr \noalign{\hrule}
height2pt
&\omit& &\omit& &\omit&
\cr
&$U_1~U_2$                     &&Drag~CDM~growth                                                       &&~\Edis\EeqnUGeneral                      & \cr
height2pt
&\omit& &\omit& &\omit&
\cr \noalign{\hrule}
height2pt
&\omit& &\omit& &\omit&
\cr
&$R~(R_t)$                     &&$b/\gamma$~momentum~(at~t)                                           &&~\Adis\AeqnSound                         & \cr
height2pt
&\omit& &\omit& &\omit&
\cr \noalign{\hrule}
height2pt
&\omit& &\omit& &\omit&
\cr
&$T$                           &&Transfer~function                                                     &&~\Edis\EeqnTparts                        & \cr
height2pt
&\omit& &\omit& &\omit&
\cr \noalign{\hrule}
height2pt
&\omit& &\omit& &\omit&
\cr
&$T_b$                         &&Baryon~$T$~drag~contrib.                                              &&~\Edis\EeqnTba~\Edis\EeqnTbi             & \cr
height2pt
&\omit& &\omit& &\omit&
\cr \noalign{\hrule}
height2pt
&\omit& &\omit& &\omit&
\cr
&$T_c$                         &&CDM~$T$~drag~contrib.                                                 &&~\Edis\EeqnTc                            & \cr
height2pt
&\omit& &\omit& &\omit&
\cr \noalign{\hrule}
height2pt
&\omit& &\omit& &\omit&
\cr
&$V_i$                         &&Velocity~in~$i$                                                       &&~\Adis\AeqnHierarchy                     & \cr
height2pt &\omit& &\omit& &\omit&
\cr \noalign{\hrule} }}}
\vskip 10pt}  
Table~1: Commonly used 
symbols.\footnote{$^{\rm\dag}$}{Fluid elements are $i=b,c,\gamma,\nu,m,r$
for the baryons, CDM, photons, neutrinos, total non-relativistic matter,
and total relativistic matter respectively.  Special epochs include
$t=*,d,eq,H$ for last scattering, the drag epoch, matter-radiation 
equality, and horizon crossing respectively.  Overdots represent 
conformal time derivatives.}
\eject
Table~1 (cont.)

\vbox{ \vskip 20pt \centerline{
\vbox{ \offinterlineskip
\halign { \vrule#
& \quad # \hfil& \vrule#
& \quad # \hfil& \vrule#
& \quad # \hfil& \vrule# \cr
\noalign{\hrule}
height2pt
&\omit& &\omit& &\omit& \cr
&$a~(a_t)$                     &&Scale~factor~(at~t)                                                   &&~\Adis\AeqnHierarchy                     & \cr
height2pt
&\omit& &\omit& &\omit&
\cr \noalign{\hrule}
height2pt
&\omit& &\omit& &\omit&
\cr
&$a_H$                         &&Horizon~crossing~$a$                                                  &&~\Bdis\BeqnaH                            & \cr
height2pt
&\omit& &\omit& &\omit&
\cr \noalign{\hrule}
height2pt
&\omit& &\omit& &\omit&
\cr
&$c_s$                         &&Photon-baryon~sound~speed                                             &&~\Adis\AeqnSound                         & \cr
height2pt
&\omit& &\omit& &\omit&
\cr \noalign{\hrule}
height2pt
&\omit& &\omit& &\omit&
\cr
&$f_\nu$                       &&Neutrino~fraction                                                     &&~\Bdis\BeqnPiZero                        & \cr
height2pt
&\omit& &\omit& &\omit&
\cr \noalign{\hrule}
height2pt
&\omit& &\omit& &\omit&
\cr
&$k$                           &&wavenumber~(Lapl.~eigenv.)                                            &&~\Adis\AeqnHierarchy                     & \cr
height2pt
&\omit& &\omit& &\omit&
\cr \noalign{\hrule}
height2pt
&\omit& &\omit& &\omit&
\cr
&$k_D$                         &&Diffusion~waven.                                                      &&~\Adis\AeqnDamp                          & \cr
height2pt
&\omit& &\omit& &\omit&
\cr \noalign{\hrule}
height2pt
&\omit& &\omit& &\omit&
\cr
&$k_{D\gamma}$                 &&CMB~damping~waven.                                                    &&~\Adis\AeqnDamp~\Ddis\DeqnKgamma         & \cr
height2pt
&\omit& &\omit& &\omit&
\cr \noalign{\hrule}
height2pt
&\omit& &\omit& &\omit&
\cr
&$k_S$                         &&Silk~damping~waven.                                                   &&~\eqnSilk~\Ddis\DeqnSilk                 & \cr
height2pt
&\omit& &\omit& &\omit&
\cr \noalign{\hrule}
height2pt
&\omit& &\omit& &\omit&
\cr
&$k_{eq}$                      &&Equality~horizon~waven.                                               &&~\Bdis\BeqnPiZero                        & \cr
height2pt
&\omit& &\omit& &\omit&
\cr \noalign{\hrule}
height2pt
&\omit& &\omit& &\omit&
\cr
&$k_{\gamma j}$              &&$j$th~photon~peak~waven.                                              &&~\eqnGammaPeak                           & \cr
height2pt
&\omit& &\omit& &\omit&
\cr \noalign{\hrule}
height2pt
&\omit& &\omit& &\omit&
\cr
&$k_{bj}$                      &&$j$th~baryon~peak~waven.                                              &&~\eqnBaryonPeak                          & \cr
height2pt
&\omit& &\omit& &\omit&
\cr \noalign{\hrule}
height2pt
&\omit& &\omit& &\omit&
\cr
&$m_\gamma$                    &&CMB~damping~steepness                                                 &&~\eqnGammaDampApp~\Ddis\DeqnSlopegamma   & \cr
height2pt
&\omit& &\omit& &\omit&
\cr \noalign{\hrule}
height2pt
&\omit& &\omit& &\omit&
\cr
&$m_S$                         &&Baryon~damping~steepness                                              &&~\eqnSilk~\Ddis\DeqnmS                   & \cr
height2pt
&\omit& &\omit& &\omit&
\cr \noalign{\hrule}
height2pt
&\omit& &\omit& &\omit&
\cr
&$q$                           &&Scaled~wavenumber                                                     &&~\Edis\EeqnBBKS                          & \cr
height2pt
&\omit& &\omit& &\omit&
\cr \noalign{\hrule}
height2pt
&\omit& &\omit& &\omit&
\cr
&$r_s$                         &&Sound~horizon                                                         &&~\Adis\AeqnSound                         & \cr
height2pt
&\omit& &\omit& &\omit&
\cr \noalign{\hrule}
height2pt
&\omit& &\omit& &\omit&
\cr
&$y$                           &&$a/a_{eq}$                                                           &&~\Edis\Eeqndeltacy                       & \cr
height2pt
&\omit& &\omit& &\omit&
\cr \noalign{\hrule}
height2pt
&\omit& &\omit& &\omit&
\cr
&$z_*$                         &&Last~scatt.~redshift                                                  &&~\Ddis\DeqnLS                            & \cr
height2pt
&\omit& &\omit& &\omit&
\cr \noalign{\hrule}
height2pt
&\omit& &\omit& &\omit&
\cr
&$z_d$                         &&Drag~redshift                                                         &&~\Ddis\DeqnZdrag                         & \cr
height2pt
&\omit& &\omit& &\omit&
\cr \noalign{\hrule}
height2pt
&\omit& &\omit& &\omit&
\cr
&$z_{eq}$                      &&Equality~redshift                                                     &&~\eqnSound                               & \cr
height2pt &\omit& &\omit& &\omit&
\cr \noalign{\hrule} }}}
\vskip 10pt}  
\eject
\section{References}

\refs Abramowitz, M. \&  Stegun, I.A. 1965 {\it 
Handbook of Mathematical Functions}, (Dover, New York) 

\refs Bardeen, J.M., Bond, J.R., Kaiser, N., \& Szalay, A.S.  1986
ApJ,  {\bf 304}, 15  (BBKS)

\refs Bond, J.R. 1996, {\it Cosmology and Large Scale Structure}, edited 
by R. Schaeffer (Elsevier Science Publishers, Netherlands) in press.

\refs Bond, J.R. \& Efstathiou, G. 1984, ApJ Lett.,  {\bf 285}, L45 

\refs Chibisov, G.V. 1972,  Astr. Zh., {\bf 49}, 72 

\refs Doroshkevich, A.G., Zel'dovich, Ya.B., \& Sunyaev, R.A. 1978,
Sov. Astron., {\bf 22}, 523 

\refs Gnedin, N.Y. \& Ostriker, J.P. 1990,  ApJ, {\bf 400}, 1 

\refs Hu, W. \& Sugiyama, N. 1995, ApJ, {\bf 444}, 489  (HSa)

\refs Hu, W. \& Sugiyama, N. 1995, Phys. Rev. D., {\bf 51}, 2599 (HSb)

\refs Hu, W., Scott, D., Sugiyama, N. \& White, M. 1995, Phys. Rev. D.,
{\bf 52}, 5498 (1995)

\refs Hu, W. \& White 1995, M., A\&A (in press, astroph-9507060)

\refs Hu, W. \& White 1996, M., ApJ (in press, astroph-9602019)

\refs Jones, B.J.T. \&  Wyse, R.F.G. 1985, A\&A, {\bf 149}, 144 


\refs Kaiser, N. 1983, MNRAS, {\bf 202} 1169 

\refs Kaiser, N. 1984, ApJ, {\bf 282}, 374 

\refs Kodama, H. \& Sasaki M. 1986, Int. J. Mod. Phys. {\bf A1}, 265 

\refs Kosowsky A. 1996, Ann. Phys., {\bf 246}, 49 

\refs Mather, J.C. \etal\ 1994, ApJ {\bf 420}, 439 

\refs Peacock, J.A. \& Dodds, S.J. 1994, MNRAS, {\bf 267}, 1020 

\refs Peebles, P.J.E. 1968, ApJ, {\bf 153}, 1 

\refs Peebles, P.J.E. 1980, {\it Large Scale Structure of the
Universe}, (Princeton University Press, Princeton) \S 12, 92.

\refs Peebles, P.J.E. 1987a, ApJ Lett., {\bf 315},  L73 

\refs Peebles, P.J.E. 1987b, Nature, {\bf 327}, 210

\refs Peebles, P.J.E. \& Yu, J.T. 1970, ApJ, {\bf 162}, 815

\refs Press, W.H. \& Vishniac, E.T. 1980, ApJ, {\bf 236}, 323 

\refs Sachs, R.K. \& Wolfe, A.M. 1967, ApJ, {\bf 147}, 73

\refs Silk, J. 1968, ApJ, {\bf 151}, 459 

\refs Sugiyama, N. 1995, ApJS, {\bf 100}, 281 

\refs Sunyaev, R.A. \& Zel'dovich, Ya. B. 1970,  Ap\&SS, {\bf 7}, 3

\refs Weinberg, S. 1972, {\it Gravitation and Cosmology} 
(Wiley, New York) p. 568.

\refs White, S.D.M.,  Navarro, J.F.,  Evrard, A.E., Frenk, C.S.  1993
Nature, {\bf 366}, 429.


\end